\documentclass[aps,prd,onecolumn,groupedaddress,nofootinbib]{revtex4}         

\usepackage[english]{babel}
\usepackage{booktabs}
\usepackage[letterpaper,top=2cm,bottom=2cm,left=3cm,right=3cm,marginparwidth=1.75cm]{geometry}

\usepackage{amsmath}
\usepackage{graphicx}
\usepackage[colorlinks=true, allcolors=blue]{hyperref}
\usepackage{xcolor}
\usepackage{multirow}

\begin{document}

\title{Constraints on Cosmic Birefringence from SPIDER, Planck, and ACT observations}

\author{Lu Yin$^{1}$}
\email{yinlu@shu.edu.cn}
\author{Shuhang Xiong$^{1}$}
\email{buffalo@shu.edu.cn}
\author{Joby Kochappan$^{2}$}
\email{jobypk@gmail.com}
\author{Bum-Hoon Lee$^{1, 3, 4}$}
\email{bhl@sogang.ac.kr}
\author{Tuhin Ghosh$^{5}$}
\email{tghosh@niser.ac.in}
\affiliation{$^{1}$Department of Physics, Shanghai University, Shanghai, 200444,  China.}
\affiliation{$^{2}$ Manipal Centre for Natural Sciences, Manipal Academy of Higher Education, Manipal, 576104, India.}
\affiliation{$^{3}$Center for Quantum Spacetime, Sogang University, Seoul 121-742, South Korea}
\affiliation{$^{4}$Department of Physics, Sogang University, Seoul 121-742, South Korea}
\affiliation{%
$^{5}$National Institute of Science Education and Research, An OCC of Homi Bhabha National Institute, Bhubaneswar, 752050, Odisha, India.}%

\date{\today}

\begin{abstract}
{The Early Dark Energy (EDE) model has been proposed as a candidate mechanism to generate cosmic birefringence through a Chern–Simons coupling between a dynamical scalar field and the cosmic microwave background (CMB) photon. Such birefringence induces a nonzero cross-correlation between the CMB $E$- and $B$-modes, providing a direct observational signature of parity violation. Recent measurements of the $EB$ and $TB$ power spectra, however, cannot yet unambiguously separate instrumental miscalibration ($\alpha$) from a true cosmic-rotation angle ($\beta$). For this reason, we perform a model-independent analysis in terms of the total effective rotation angle $\alpha+\beta$.
We analyze the latest $EB$ and $TB$ measurements from the SPIDER, \textit{Planck}, and ACT experiments and derive constraints on the Chern–Simons coupling constant $gM_{Pl}$ and on the polarization rotation angle $\alpha+\beta$. We find that the coupling $gM_{Pl}$ provides reasonable fits to the SPIDER, \textit{Planck} and ACT measurements. The fits for $\alpha+\beta$ prefer a value larger than zero: when combined, \textit{Planck}+ACT yield a detection significance of approximately 7$\sigma$.
We also find that ACT data alone do not provide sufficiently tight constraints on either $gM_{Pl}$ or $\alpha+\beta$, whereas the combination \textit{Planck}+ACT improves the statistical significance of ACT’s high-$\ell$ results and leads to a better PTE for those measurements. }

\end{abstract}

\maketitle

\section{Introduction}

The discovery of the Cosmic Microwave Background (CMB) was a turning point in twentieth-century physics, providing decisive evidence for the hot Big Bang picture and establishing the $\Lambda$CDM model as the standard cosmological framework \cite{Komatsu:2014ioa, Planck:2018vyg, SPT-3G:2021eoc, BICEP:2021xfz, SPIDER:2021ncy}. Although this model matches most observations with remarkable accuracy, recent advances in precision cosmological observation have begun to expose possible cracks in the paradigm \cite{Abdalla:2022yfr}. Two anomalies in particular motivate the present work: the hints of cosmic birefringence \cite{Feng:2004mq, Komatsu:2022nvu} and the Hubble constant discrepancy \cite{Bernal:2016gxb}.

Cosmic birefringence refers to a rotation in the linear polarization of CMB photons as they travel across cosmic distances \cite{Carroll:1989vb, Carroll:1991zs, Harari:1992ea}. Analyses of Planck data suggest a non-zero rotation angle $\beta$ with a significance of about $3.6 \sigma$, corresponding to $\beta={0.342^{\circ}}^{+0.094^\circ}_{-0.091^\circ}$ (68$\%$ C.L.) \cite{Minami:2020odp, Eskilt:2022cff}. The phenomenon reveals itself through correlations between the $E$ and $B$ polarization modes \cite{Lue:1998mq}. In standard parity-conserving physics, $E_{\ell m}$ and $B_{\ell m}$ have well-defined transformation rules under inversion, which guarantee that their auto-spectra remain unchanged while their cross-spectrum $C_\ell^{EB}$ flips sign. A non-zero $EB$ signal therefore points directly to parity violation\cite{Naokawa:2023upt, Namikawa:2023zux, Ferreira:2023jbu, Yin:2023srb, Greco:2024oie, Namikawa:2024dgj, Sullivan:2025btc, LiteBIRD:2025yfb, Lonappan:2025hwz, Namikawa:2025sft, Ballardini:2025apf, Namikawa:2025doa, Diego-Palazuelos:2025dmh}, even for CPT violation \cite{Feng:2006dp, Li:2006ss, Li:2008tma, Xia:2008si, Xia:2007qs, Xia:2009ah, Li:2014oia} and the anisotropic effect \cite{Li:2013vga, Luongo:2021nqh, Krishnan:2021dyb, Namikawa:2024sax}. One compelling possibility is the existence of axion-like fields that couple to the electromagnetic tensor via an interaction $g\phi F_{\mu\nu}\tilde{F}^{\mu\nu}$ \cite{Choi:2021aze, Nakatsuka:2022epj, Murai:2022zur}. Such term would rotate polarization directions and generate $EB$ correlations, while simultaneously leaking power from $E$ into $B$ modes. {At present, there are already many detectors that have published their observation for $EB$ correlation. For example, {such as} POLARBEAR \cite{POLARBEAR:2019snn}, ACT \cite{ACT:2025fju}, SPT \cite{SPT-3G:2021eoc}, and SPIDER \cite{SPIDER:2021ncy}. } Upcoming polarization experiments, including the Simons Observatory \cite{SimonsObservatory:2018koc}, {AliCPT \cite{Gao:2017cra, Li:2017drr}}, and LiteBIRD \cite{LiteBIRD:2022cnt}, are expected to probe these effects with much higher sensitivity. In particular, LiteBIRD is forecast to detect not only a potential $EB$ signal but also the secondary $BB$ component sourced by birefringence.

Parallelly, the Hubble tension is another pressing challenge faced in front of cosmology. This tension is the
mismatch between the Hubble constant inferred from early-Universe probes such as the CMB and Baryon Acoustic Oscillations (BAO), and the values obtained from local measurements via the distance ladder \cite{Bernal:2016gxb, Riess:2021jrx}. The discrepancy at the $4$–$5\sigma$ level and cannot be ignored nowadays \cite{Escamilla:2024ahl, Du:2024pai, Li:2024qso,Cai:2025mas,Li:2025nnk, Qiu:2024sdd, Feng:2025mlo, Smith:2025icl, Lee:2025yvn, Piras:2025eip, Li:2025owk}. Theoretical explanations have increasingly focused on modifications of the cosmic expansion history. Among the few proposals able to alleviate the tension without creating fresh inconsistencies, Early Dark Energy (EDE) scenarios have emerged as particularly promising. The EDE model posits the presence of a transient dark energy component in the pre-recombination Universe, which alters the sound horizon at last scattering and thereby shifts the inferred value of $H_0$. EDE has been extensively tested against observational data \cite{Yin:2020dwl, Colgain:2021pmf, Krishnan:2021dyb, Krishnan:2021jmh, Akarsu:2022lhx, Murai:2022zur,Poulin:2018cxd,Herold:2023vzx,Efstathiou:2023fbn,Simon:2024jmu,Lin:2025gne} with mixed results, some articles find that EDE does not fit the cosmic birefringence observations \cite{Eskilt:2023nxm}, while others find that EDE is fully consistent with the cosmic birefringence measurements as well as the $H_0$ measurements from the SH0ES experiment \cite{Kochappan:2024jyf}.
{In this paper, we will investigate the cosmic birefringence effect resulting from the introduction of a scalar-photon coupling within the EDE framework.} Such a case creates a natural link between the physics of the Hubble tension and that of Cosmic Birefringence. Our central question is under multi CMB detectors at present whether the 
$EB$ observation gives a consistent Chern-Simons coupling constant? Are the miscalibration and rotation angle ($\alpha+\beta$) consistent with different $EB$ data?
To address this, we selected SPIDER, Planck, and ACT data to constrain the constant $gM_{Pl}$ and model-independent angle $\alpha+\beta$, separately.

The structure of the paper is as follows. Section \ref{sec:2}  reviews the EDE model we examine, outlines the Boltzmann equations, and explains how they are altered by scalar–photon couplings. 
Section \ref{sec:3} presents our results; the constraint results of $gM_{Pl}$, $\alpha+\beta$, and $\chi^2$ in different datasets will be discussed. 
We summarize the results in Section \ref{sec:4}.

\section{Cosmic Birefringence from the Early Dark Energy model }
\label{sec:2}

By introducing a Chern–Simons interaction, the Lagrangian describing the coupling between a pseudoscalar field and the electromagnetic field can be expressed as \cite{Murai:2022zur}
\begin{equation}
    \mathcal{L}=-\frac{1}{2}\left(\partial_\mu \phi\right)^2-V(\phi)-\frac{1}{4} F_{\mu \nu} F^{\mu \nu}-\frac{1}{4} g \phi F_{\mu \nu} \tilde{F}^{\mu \nu},
\end{equation}
where $\phi$ represents a pseudoscalar field with a standard kinetic term and a potential $V(\phi)$. The constant $gM_{Pl}$, with mass dimension $-1$, determines the strength of the Chern–Simons coupling. Here, $F_{\mu\nu}$ is the electromagnetic field strength tensor, and $\tilde{F}_{\mu\nu}$ denotes its dual counterpart.

{Early dark energy (EDE) refers to a cosmological component that contributes a significant fraction of the total energy density during the matter-radiation equality epoch. It has been proposed as a potential solution to the Hubble tension by altering the pre-recombination expansion rate \cite{Caldwell:2003vp, Smith:2019ihp, Berghaus:2019cls, Alexander:2019rsc, Chudaykin:2020acu, Agrawal:2019lmo, Niedermann:2019olb, Freese:2004vs, Ye:2020btb, Akarsu:2019hmw, Lin:2019qug, Yin:2020dwl, Braglia:2020bym}.
The potential of the EDE in general is written as 
\begin{equation}
    V(\phi)=\Lambda^4(1-\cos(\phi/f))^n,
    \label{eq:pot}
\end{equation}
where $n$ {is a phenomenological parameter. In this work, we focus on the EDE model with $n$=3.} 
In such models, the energy density becomes relevant around, leading to a reduction of the comoving sound horizon prior to photon decoupling. And the Eq.(\ref{eq:pot}) provided a pseudoscalar potential and this EDE can be the birefringence material in our Universe.

}

The presence of the Chern–Simons term in the Lagrangian alters the propagation properties of photons, leading to a modified dispersion relation \cite{Carroll:1989vb, Carroll:1991zs, Harari:1992ea}
\begin{equation}
\omega_{\pm} \simeq k \mp \frac{g}{2} \left( \frac{\partial \phi}{\partial t} + \frac{\mathbf{k}}{k} \cdot \nabla \phi \right) = k \mp \frac{g}{2} \frac{d\phi}{dt},
\end{equation}
where the symbols $+$ and $-$ indicate the right- and left-handed circular polarization modes of the photon, and $\omega_{\pm}$ represents the angular frequency associated with each helicity. We adopt a right-handed coordinate system with the $z$-axis aligned along the photon’s direction of motion.

The rotation experienced by the plane of linear polarization is directly determined by the helicity-dependent dispersion relation. In the regime where the photon frequency $\omega_\pm$ is much larger than the rate of change of the scalar field $\phi$, the WKB approximation can be applied. Within this framework, the cumulative rotation angle of the polarization from an initial time $t$ to the present time $t_0$ is given by
\begin{equation}
\label{beta}
\beta(t) = -\frac{1}{2} \int_t^{t_0} d\tilde{t} , (\omega_+ - \omega_-) = \frac{g}{2} \left[ \phi(t_0) - \phi(t) \right].
\end{equation}

Under the conventions adopted here, a positive rotation angle, $\beta > 0$, corresponds to a clockwise rotation of the linear polarization as seen on the sky, with the $z$-axis defined along the observer’s line of sight. Our choice of polarization angle follows the same conventions as those in \cite{Komatsu:2022nvu}.

The evolution of the CMB polarization, including the effect of cosmic birefringence, can be described by the modified Boltzmann equation:

\begin{equation}
{ }_{\pm 2} \Delta_P^{\prime}+i q \mu_{\pm 2} \Delta_P=  \tau^{\prime}\left[-{ }_{\pm 2} \Delta_P+\sqrt{\frac{6 \pi}{5}}{ }_{\pm 2} Y_2^0(\mu) \Pi(\eta, q)\right]  \pm 2 i \beta^{\prime}{ }_{\pm 2} \Delta_P,
\end{equation}
where $\eta$ denotes the conformal time, $q$ is the Fourier space wave number, and $\mu$ represents the cosine of the angle between the photon propagation direction and the wave vector. The symbols ${ }{\pm 2} Y_\ell^m$ correspond to the spin-2 spherical harmonics, and $\Pi(\eta, q)$ defines the polarization source function. The quantity ${ }_{\pm 2} \Delta_P$ represents the Fourier transform of $Q \pm i U$, with $Q$ and $U$ being the standard Stokes parameters describing the linear polarization of the radiation.

To express the polarization in terms of the spin-2 harmonics, ${ }_{\pm 2} \Delta_P$ can be expanded as

\begin{equation}
{ }_{\pm 2} \Delta_P(\eta, q, \mu) \equiv \sum_\ell i^{-\ell} \sqrt{4 \pi(2 l+1)}{ }_{\pm 2} \Delta_{P, l}(\eta, q)_{\pm 2} Y_\ell^0(\mu).
\end{equation}

In general, the rotation angle $\beta$ varies with conformal time, and the solution of the Boltzmann equation incorporating $\beta(\eta)$ can be written as

\begin{equation}
\begin{array}{rl}
{ }_{\pm 2} \Delta_{P, l}\left(\eta_0, q\right)=-\frac{3}{4} \sqrt{\frac{(l+2) !}{(l-2) !}} \int_0^{\eta_0} \mathrm{~d} \eta \tau^{\prime} e^{-\tau(\eta)} \Pi(\eta, q) \times \frac{j_\ell(x)}{x^2} e^{\pm 2 i \beta(\eta)} , 
\end{array}
\label{Boltzmann}
\end{equation}
where $x = q(\eta_0 - \eta)$ and $j_\ell(x)$ is the spherical Bessel function of order $\ell$. This formulation explicitly incorporates the time-dependent rotation of the polarization plane into the Fourier-space multipoles of the polarization.

The angular power spectra of CMB polarization can be expressed as

\begin{equation}
C_\ell^{X Y}=4 \pi \int \mathrm{d}(\ln q) \mathcal{P}_s(q) \Delta_{X, l}(q) \Delta_{Y, l}(q) ,
\end{equation}
where $\mathcal{P}_s(q)$ denotes the primordial scalar curvature power spectrum, and $X,Y$ indicate the $E$ or $B$ polarization modes. The multipole coefficients $\Delta_{E,\ell}(q)$ and $\Delta_{B,\ell}(q)$ can be directly obtained from Eq.~\ref{Boltzmann} through the relation

\begin{equation}
\Delta_{E, l}(q) \pm i \Delta_{B, l}(q) \equiv-{}_{\pm 2}\Delta_{P, l}\left(\eta_0, q\right).
\end{equation}

Further discussions on the Boltzmann treatment of $TB$ and $EB$ correlations can be found in Refs.~\cite{Finelli:2008jv, Galaverni:2023zhv}.
In the special case where the rotation angle $\beta$ is constant, the observed $E$ and $B$ mode fluctuations can be expressed as

\begin{equation}
\Delta_{E, \ell} \pm i \Delta_{B, \ell}=e^{\pm 2 i \beta}\left(\tilde{\Delta}_{E, \ell} \pm i \tilde{\Delta}_{B, \ell}\right),
\end{equation}
where $\tilde{\Delta}_{E, \ell}$ and $\tilde{\Delta}_{B, \ell}$ represent the fluctuations in the absence of Cosmic Birefringence. In this framework, any non-zero $EB$ {and $TB$} correlations reflect parity-violating effects \cite{Lue:1998mq}.

\begin{equation}
\begin{array}{l}
C_\ell^{E E}=\cos ^2(2 \beta) \tilde{C}_\ell^{E E}+\sin ^2(2 \beta) \tilde{C}_\ell^{B B}  ,
\end{array}
\end{equation}
\begin{equation}
	\begin{array}{l}
		C_\ell^{B B}=\cos ^2(2 \beta) \tilde{C}_\ell^{B B}+\sin ^2(2 \beta) \tilde{C}_\ell^{E E}  ,
	\end{array}
\end{equation}
\begin{equation}
	\begin{array}{l}
		C_\ell^{E B}=\frac{1}{2} \sin (4 \beta)\left(\tilde{C}_\ell^{E E}-\tilde{C}_\ell^{B B}\right),
	\end{array}
 \label{eq:cleb}
\end{equation}
  \begin{equation}
	\begin{array}{l}
		C_\ell^{T B}=\sin (2 \beta)\tilde{C}_\ell^{T E},
	\end{array}
 \label{eq:tb}
\end{equation}
where $\tilde{C}_\ell$ represents the polarization spectra in the absence of any {Cosmic Birefringence}. When $\beta$ is set to zero, the { $TE$,} $EE$ and $BB$ spectra reduce to their standard values without photon-scalar interactions, and the $EB${, $TB$} cross spectrum vanishes. In cases where the $BB$ spectrum is much smaller than $EE$, the $EB$ correlation can be approximated as
$
C_\ell^{EB} \simeq \tan(2\beta) C_\ell^{EE}.
$
This formulation clearly illustrates how a constant birefringence angle mixes the $E$ and $B$ modes and induces a nonzero $EB$ { and $TB$} correlation.

In general, the { potential $\phi$ varies over time from} the dynamical evolution of the scalar. In such cases, an analytical treatment of the rotation angle $\beta$ and the Boltzmann equations must be integrated numerically. To accomplish this, we employ the publicly available \texttt{CLASS\_{EDE}} extension \cite{Hill:2020osr} of the \texttt{CLASS} code \cite{Lesgourgues:2011re, Blas:2011rf} to compute the polarization power spectra $EE$, $BB$, $TE$, $TB$ and $EB$ 
We will analysis concentrate on the EDE with different datasets from SPIDER, Planck, and ACT.

\section{Fitting cosmic birefringence with different CMB datasets}
\label{sec:3}

In this Section, we use $EB$ {and $TB$} observation data from SPIDER, Planck, and ACT {experiments} to fit the Chern-Simons coupling constant parameter $ g$ {  based on the EDE model. Meanwhile, we also constrain the model-independent angle $\alpha+\beta$, which is the addition of miscalibration and cosmic birefringence rotation angle}.  
SPIDER is a balloon-borne CMB instrument designed primarily to measure the polarization of the {CMB} on degree angular scales, with Galactic synchrotron emission being negligible within its observational frequency bands. SPIDER provides observations at very low multipole moments $(\ell < 250)$, and currently offers 9 $EB$ data points. We use its $EB$ Combined data results\cite{SPIDER:2021ncy}.  
Meanwhile, we also considered the Planck observation. Planck is one of the most important CMB detections, providing precise $TT$, $TE$, and $EE$ polarization. We employ the $EB$ data analyzed by {Eskilt et al.} \cite{Eskilt:2022cff}, which consists of 72 data points covering the multipole range of $\ell$ between 50 and 1500. 
The third kind of $EB$ observation we considered in this paper is the {Atacama Cosmology Telescope(ACT).}
ACT is a new-generation ground-based CMB experiment, which has garnered significant attention for its performance in high-$\ell$ measurements. 
We use the latest released $EB$ data\cite{ACT:2025fju}, which includes 38 data points covering the multipole range from 600 to 3400. {These high-$l$ $TB$ data will be discussed for} its constraints on the Chern-Simons coupling constant parameter {$gM_{Pl}$}. 

We adopt the values of other parameters in the EDE model based on the best-fit results from the Base and Base+SH0ES datasets presented in {Eskilt} et al\cite{ Eskilt:2023nxm}. The Base dataset includes the Planck power spectra of temperature and polarization  \cite{Planck:2019nip}, and the Baryon Oscillation Spectroscopic Survey (BOSS) Data Release 12 \cite{BOSS:2016wmc}.
For the Base+SH0ES dataset, the local SH0ES measurement of $H_0$ is also considered \cite{Riess:2021jrx} based on Planck data.
The 9 basic EDE model parameters' numerical results are summarized in Table \ref{tab:base}. In addition, we also considered testing $gM_{Pl}$ with the best-fit parameters from the constraints with Base+SH0ES plus Planck $EB$ and Lensing data, { as we show in the previous work\cite{Kochappan:2024jyf}.  We give this group of data a short name as BSL, and show the result in} Table \ref{tab:base}.
Cosmic birefringence is treated as a secondary effect under these constrained parameter configurations.

\begin{table}[]
    \centering
\begin{tabular}{c|ccc }
\hline & Base & Base+SH0ES& BSL \\
\hline$f_{\mathrm{EDE}}$ & 0.0872 & 0.1271 & { 0.195} \\
$\log _{10} z_c$ & 3.560 & 3.563 & { 3.4752} \\
$\theta_i$ & 2.749 & 2.768 & {  1.89 } \\
\hline $100 \omega_{\mathrm{b}}$ & 2.265 & 2.278  & { 2.272}\\
$\omega_{\mathrm{CDM}}$ & 0.1282 & 0.1324  & {  0.1558 }\\
$100 \theta_s$ & 1.041 & 1.041 & {  1.0404} \\
$\ln \left(10^{10} A_s\right)$ & 3.063 & 3.071 &  {3.153} \\
$n_s$ & 0.983 & 0.992 & { 0.9887} \\
$\tau$ & 0.0562 & 0.0568 & { 0.0679} \\
\hline
\end{tabular}
    \caption{The values of the nine fixed parameters in the analysis, set to the best-fit results excluding $gM_{Pl}$, which are mentioned in \cite{Eskilt:2023nxm, Kochappan:2024jyf}.
    Base and Base+SH0ES mean that the other EDE parameters are fixed under the best-fit result from Planck+BOSS data, and Planck+BOSS+SH0ES $H_0$ data from \cite{Eskilt:2023nxm}, respectively. BSL refers to the best-fit results from our previous work \cite{Kochappan:2024jyf} include Planck $EB$ and Lensing in addition to Base+SH0ES. We give this dataset a short name, BSL, in the next.
    }
    \label{tab:base}
\end{table}

We will use the aforementioned constraint results to compute the best-fit value of {$gM_{Pl}$} and the corresponding $ \chi^2 $ statistic.  
The $\chi^2 $ formula is defined as follows:  
\begin{equation}
\chi^2=\sum_{i=1}^n \frac{\left(T_i-O_i\right)^2}{\sigma_i^2},
\end{equation}
where $T_i$ is the theoretical prediction, and $O_i$ ($\sigma_i$) is the observational value  {of the CMB power spectra, respectively}.

Given the varying numbers of observational data points across SPIDER, Planck, and ACT datasets, we will use the probability to exceed (PTE) to evaluate the quality of the fit to the data. The PTE value represents the probability of finding a $\chi^2$ value greater than the one obtained, given the number of degrees of freedom. A small PTE means that the $\chi^2$ is too high for the degrees of freedom and is indicative of a poor quality fit. For reference, for large degrees of freedom, the $\chi^2$ distribution approaches a normal distribution and a PTE smaller than 0.32 corresponds to a $>1-\sigma$ tension with the data. On the other hand, a value of PTE that is very close to 1, such as 0.99, indicate overfitting.
 {The number of degrees of freedom, $\nu$, is given by, $\nu=N-k$, where $N$ is the number of observational data points, and $k$ is the number of fitted parameters}. In this fit, only $gM_{Pl}$ {  and $\alpha+\beta$ is constrained separately, so the number of parameters is $k = 1$.}

\begin{table}[]
    \centering
\begin{tabular}{|c|c|c|c|c|}
\hline & & Base & Base+SH0ES & { BSL} \\
\hline 
\multirow{2}{*}{SPIDER-$EB$} &{$gM_{Pl}$} & $1.15$ & $-3.47$ &  $0.155$ \\
\cline{2-5} 
                    & $\chi^2$      & $3.64$ &$3.48$ & { $3.45$}\\
\cline{2-5}
                     & $\nu$      & $8$ &$ 8$ & { $ 8 $}\\
\cline{2-5}
\cline{2-5}
     & PTE    & $0.888$ &$0.901$ & $0.903$\\ 
\hline 
\hline 
\multirow{2}{*}{Planck-$EB$} & {$gM_{Pl}$} & $0.54$ & $0.62$ & { $0.087$} \\
\cline{2-5}
   & $\chi^2$ & $77.62$ &$103.05$ & { $69.07$}\\
\cline{2-5}
                     & $\nu$      & $ 71$ &$ 71$ & { $ 71$}\\
\cline{2-5}
\cline{2-5}
     & PTE    & $0.276$ &$0.0077$ & { $0.542 $}\\
\hline 
\hline 
\multirow{2}{*}{ACT-$EB$} & {$gM_{Pl}$} & $0.05$ & $0.002$ & { $0.127$}\\
\cline{2-5}
 & $\chi^2$  & $85.98$ &$86.96$  & $50.10$ \\
 \cline{2-5}
                     & $\nu$      & $37 $ &$ 37$ & { $ 37$}\\
\cline{2-5}
\cline{2-5}
     & PTE    & $ 1\times10^{-6} $ &$6.68\times10^{-6} $ & { $ 0.0737 $}\\
\hline
\hline 
\multirow{2}{*}{ACT-$TB$} & {$gM_{Pl}$} & $-0.003$ & $-0.005$ & { $-0.153$}\\
\cline{2-5}
 & $\chi^2$  & $46.92$ &$46.88$ & { $44.65$}\\
\cline{2-5}
                     & $\nu$      & $ 37$ &$ 37$ & { $ 37$}\\
\cline{2-5}

\cline{2-5}
     & PTE    & $0.127 $ &$ 0.128 $ & { $ 0.181 $}\\
\hline
\end{tabular}
    \caption{ Best fitting results for the Chern-Simons coupling constant {$gM_{Pl}$}, $\chi^2$,  and Probability to exceed (PTE) from the dataset of SPIDER-$EB$, Planck-$EB$, ACT-$EB$, and ACT-$TB$, respectively.
}
    \label{tab:chi2-single}
\end{table}

\begin{table}[]
\centering
\begin{tabular}{c|c|c|c }
\hline & $\alpha+\beta$ & $\chi^2$ & {PTE} \\
\hline 
SPIDER-$EB$ & ${0.35^{\circ}}\pm{0.69^{\circ}}$  & 3.51 & $0.8988$ \\

Planck-$EB$ & $0.29^{\circ}\pm{0.03^{\circ}}$ & 65.84  & 0.6509 \\

ACT-$EB$ &  ${0.19^{\circ}}^{+0.03^{\circ}}_{-0.04^{\circ}}$ & 109.27  & 0.2257  \\

ACT-$TB$ &  $0.19^{\circ}\pm{0.11^{\circ}}$ & 83.39  & 0.8698 \\

\hline 

SPIDER+Planck & $0.29^{\circ}\pm{0.03^{\circ}}$  & 67.75  & 0.5539 \\

SPIDER+ACT {($EB$)} & $0.19^{\circ}\pm{0.04^{\circ}}$ & 112.83  & 0.3559  \\

Planck+ACT {($EB$)} & $0.21^{\circ}\pm{0.03^{\circ}}$ & 126.73  & 0.4900  \\

SPIDER+Planck+ACT {($EB$)} & $0.22^{\circ}\pm{0.03^{\circ}}$  & 135.36  & 0.2685 \\
\hline
\end{tabular}
    \caption{Best fitting results for a constant $\alpha+\beta$ and the corresponding $\chi^2$ and PTE values, using different combinations of the CMB $EB$ and $TB$ power spectra from SPIDER, Planck, and ACT.}
    \label{tab:alpha+beta}
\end{table}

\begin{figure}
\centering
\includegraphics[width=0.7\linewidth]{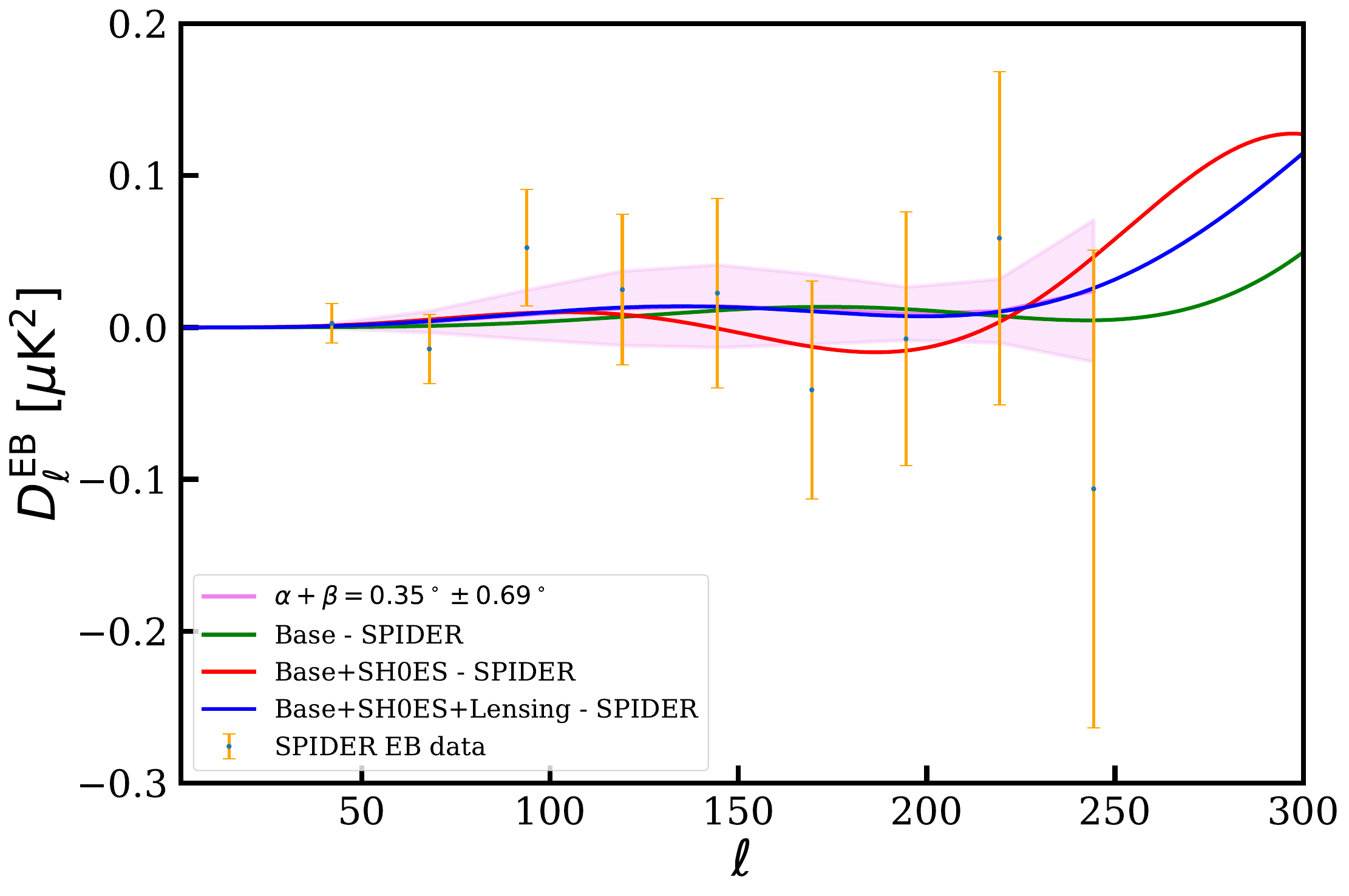}
\caption{\label{fig:spider}  Best fitting results of $gM_{Pl}$ and a constant rotation angle $\alpha+\beta$ using the CMB $EB$ power spectrum from SPIDER. Similar to Table \ref{tab:chi2-single}, Base, Base+SH0ES, and BSL, refer to fixing the remaining nine EDE parameters to the best-fit results from \cite{Eskilt:2023nxm, Kochappan:2024jyf}. The constraint result of $\alpha+\beta$ from SPIDER-$EB$ is $0.35^\circ\pm{0.69^\circ}$}.
\end{figure}

We now incorporate the {$EB$} data from SPIDER to constrain the best-fit value of the Chern-Simons coupling constant $gM_{Pl}$ and compute the corresponding $\chi^2$. Figure \ref{fig:spider} shows the CMB $ D_l^{EB} $ spectra for different values of $gM_{Pl}$ given by SPIDER-EB under the Base, Base+SH0ES,  and BSL scenarios. 
The transform function from $C_l$ to $D_l$ can be given as 
\begin{equation}
D_l=\frac{l(l+1)}{2 \pi} C_l.
\end{equation}

In the Base dataset case, the best-fit value of {$gM_{Pl}$} from SPIDER $EB$ data is 1.15, and the $\chi^2$ is 3.64. In the case of Base+SH0ES dataset, the {$gM_{Pl}$} best-fit result is -3.47, yielding a $\chi^2$ is 3.48. {  In BSL dataset, the {$gM_{Pl}$} is 0.155 and $\chi^2$ is 3.45. These results are summarized in Table \ref{tab:chi2-single}.}
The number of degrees of freedom is $\nu = N - k $ with $N = 9$ (number of data points) and $k = 1$ (number of fitted parameters). The PTE values are $0.888$, $0.901$, and $0.903$ for the Base, Base+SH0ES, and BSL datasets, respectively. The results are also shown in Table \ref{tab:chi2-single}.
 These values of PTE result suggest  that $gM_{Pl}$ provides a good fit to the SPIDER data.

\begin{figure}
\centering
\includegraphics[width=0.7\linewidth]{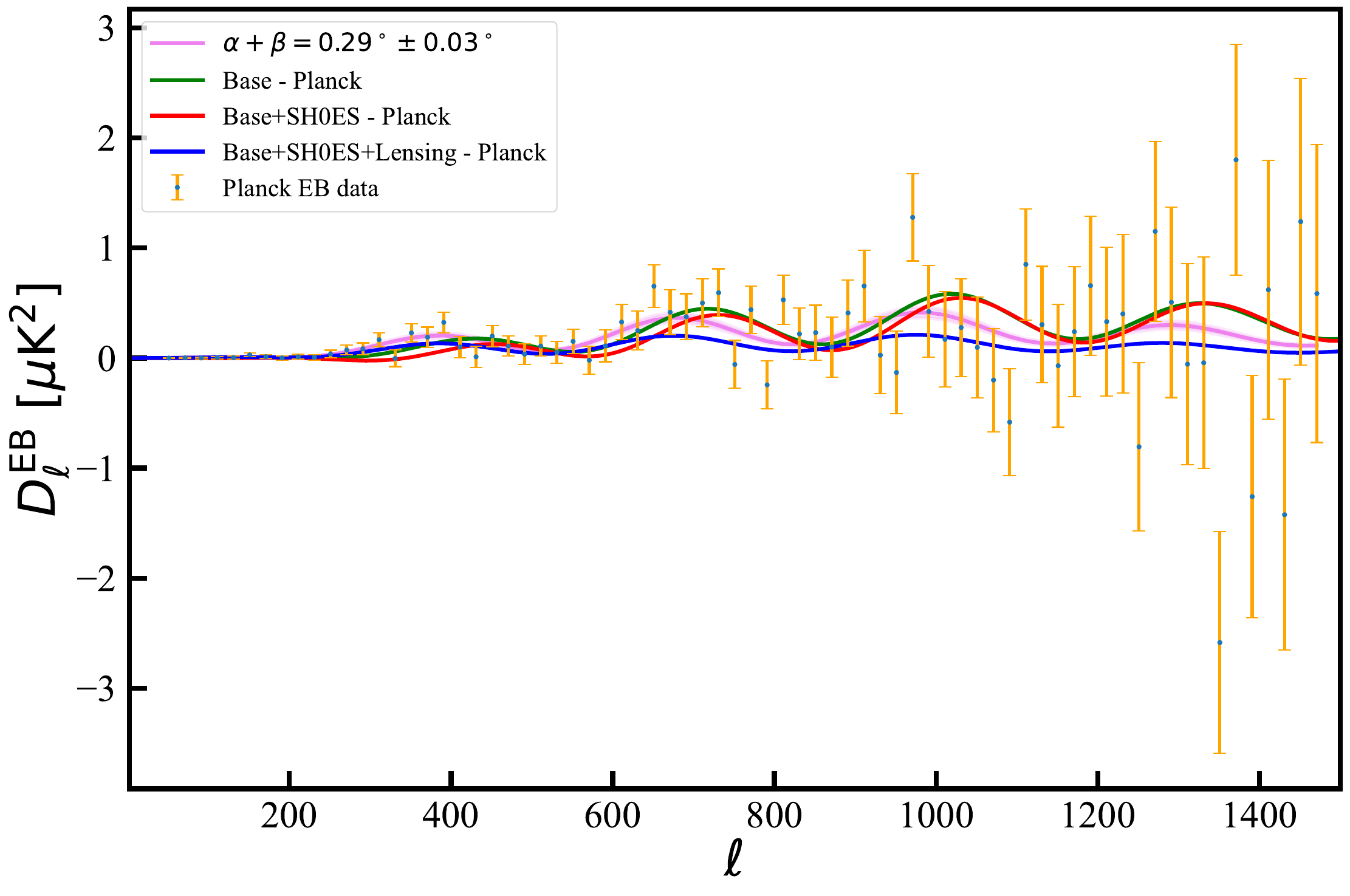}
\caption{\label{fig:planck} {The CMB power spectrum of $EB$ mode, that $EB$ data from Planck 2018. 
The other EDE parameters were fixed by the Base, Base+SH0ES, and BSL datasets. The Chern-Simons constant $gM_{Pl}$ and the minimum of $\chi^2$ are obtained by Planck-$EB$. The constraint result of $\alpha+\beta$ from Planck-$EB$ is $0.29^\circ\pm{0.03^\circ}$.}}
\end{figure}

With considering Planck, we simultaneously fitted $gM_{Pl}$ under both the Base, Base+SH0ES, and BSL scenarios and computed the corresponding $\chi^2$ values. A comparison between the theoretical $D_l^{EB}$ and observational data is shown in Fig. \ref{fig:planck}.  
 These results are consistent with those reported by the reference \cite{Eskilt:2023nxm}.
In the Base data case, we show ${gM_{Pl}} = 0.54$ with $\chi^2 = 77.62$. For the Base+SH0ES dataset, the best-fit value is ${gM_{Pl}}= 0.62$ with $\chi^2 = 103.05$. { In BSL dataset, {$gM_{Pl}$} get 0.087 with the $\chi^2=$69.07. Taking into account the 72 number of data points and 1 additional parameter, the PTE values are 0.276, 0.0077, and 0.542 for the Base, Base+SH0ES, and BSL cases, respectively.}
 {This means that $gM_{Pl}$ can fit the Planck $EB$ data when using the background cosmological parameters from the Base and BSL sets, but not with the Base+SH0ES set, which indicates a $\approx$3$\sigma$ tension.}

\begin{figure}
\centering
\begin{tabular}{cc}
\includegraphics[width=0.5\linewidth]{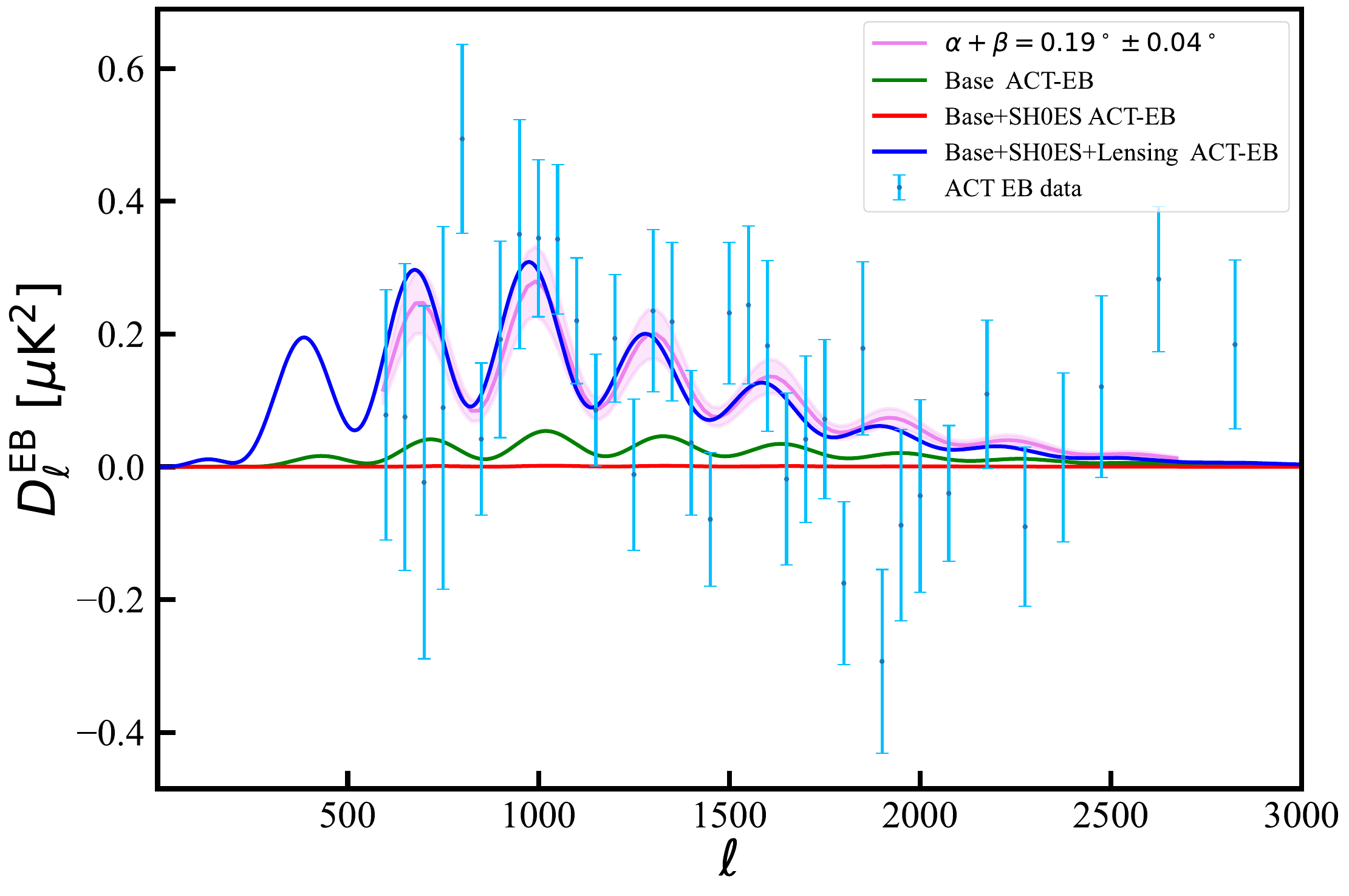}&
\includegraphics[width=0.5\linewidth]{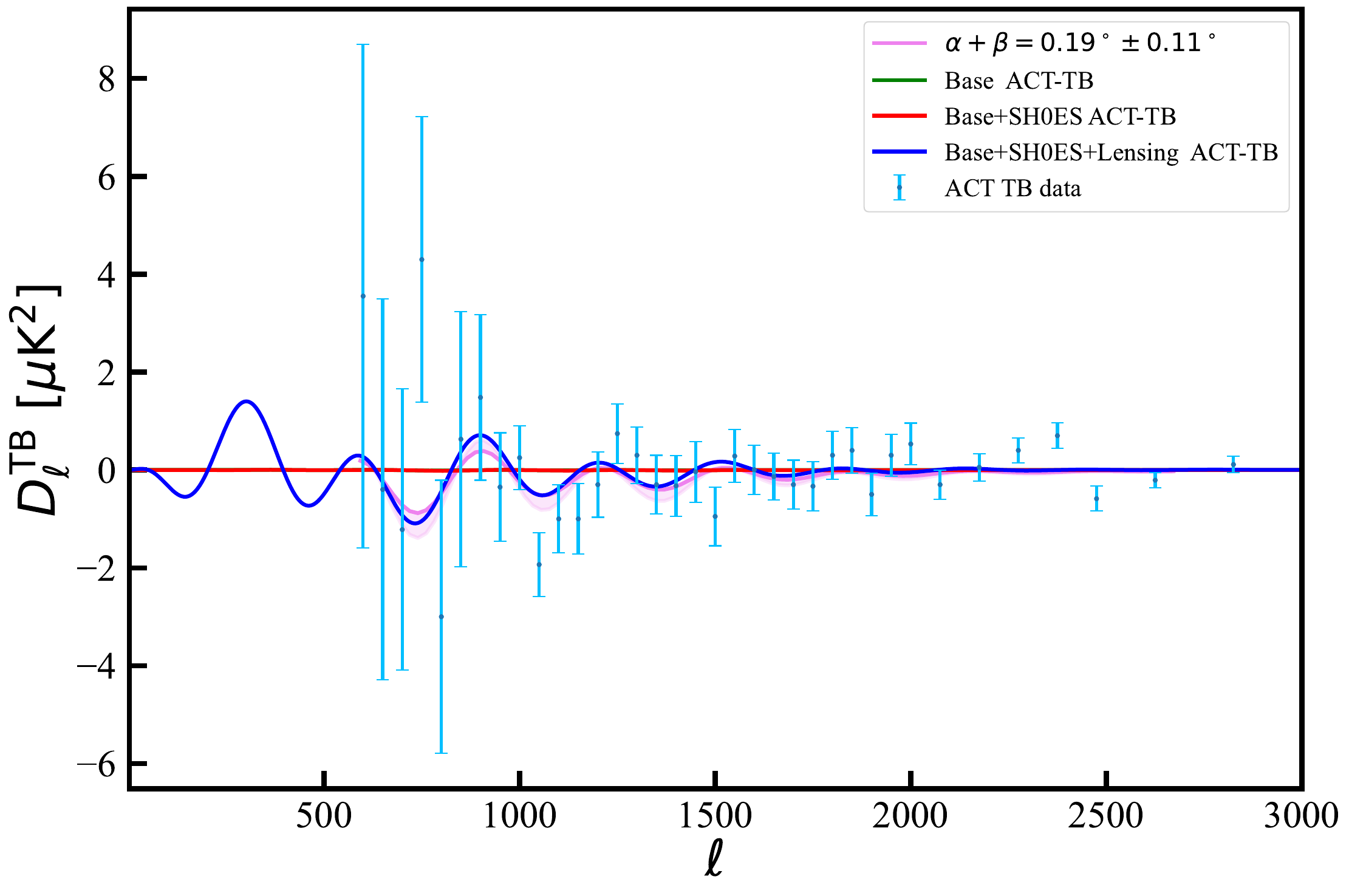}\\
(a)  & (b)   \\
    \end{tabular}
\caption{\label{fig:act} {The CMB power spectra of $EB$ mode with the ACT-$EB$ data (a), and the CMB $TB$ mode with ACT-$TB$ data (b). The green, red, and blue lines from the different values of $gM_{Pl}$ constraints by ACT and with other parameters fixed in Base, Base+SH0ES, and BSL datasets, respectively.}}
\end{figure}

We also considered constraints on $gM_{Pl}$ from the recently released $EB$ and $TB$ data from the ACT telescope. Under the Base dataset, ACT $EB$ gives ${gM_{Pl}} = 0.05$, while $TB$ gives ${gM_{Pl}} = -0.003$. For the Base+SH0ES dataset, ACT $EB$ get {$gM_{Pl}$} equal to 0.002, while $TB$ gives ${gM_{Pl}} = -0.005$. BSL case gives $gM_{Pl}$ value as 0.127 and -0.153 in $EB$ and $TB$ observation, respectively. 
The small values of the Chern-Simons coupling constant from ACT hint that the rotation angle of cosmic birefringence may be smaller than we expected from Planck.
The results are presented in Fig. \ref{fig:act} and Table \ref{tab:chi2-single}. The PTE values for the ACT $EB$ data indicate that  $gM_{Pl}$ does not fit the ACT $EB$ data in the Base and Base+SH0ES case, while providing a reasonable fit in the BSL case. Finally, we note that $gM_{Pl}$ also provides reasonable fits to the ACT $TB$ data in all three cases.
{Moreover, since $gM_{Pl}$ back to zero implies no birefringence, the  $gM_{Pl}$ with different sides of zero from the best-fit result of ACT $EB$ and $TB$ suggests that ACT data did not provide consistent constraints on $gM_{Pl}$.}

\begin{figure}
\centering   
\begin{tabular}{cc}
\includegraphics[width=0.5\linewidth]{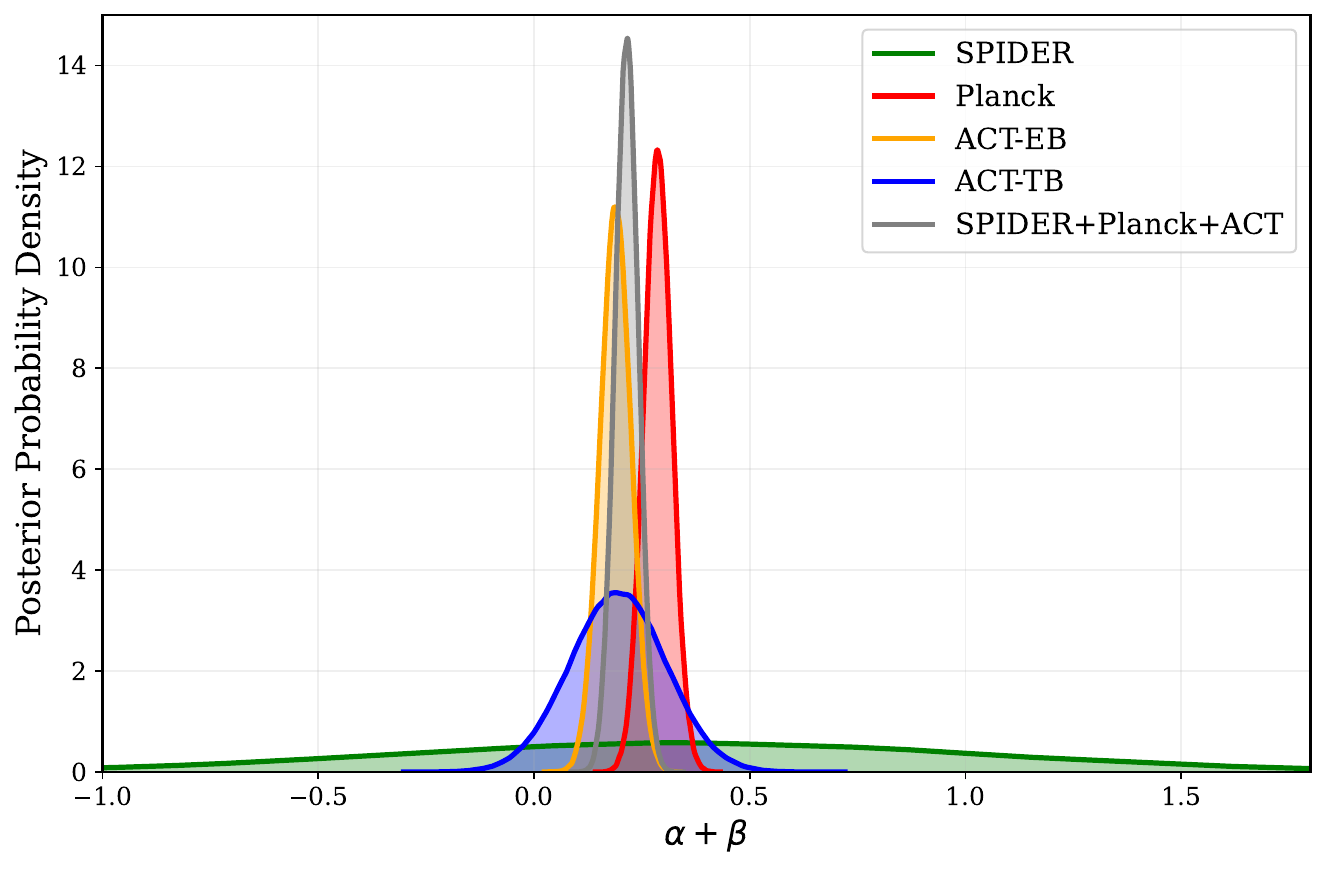}&
\includegraphics[width=0.5\linewidth]{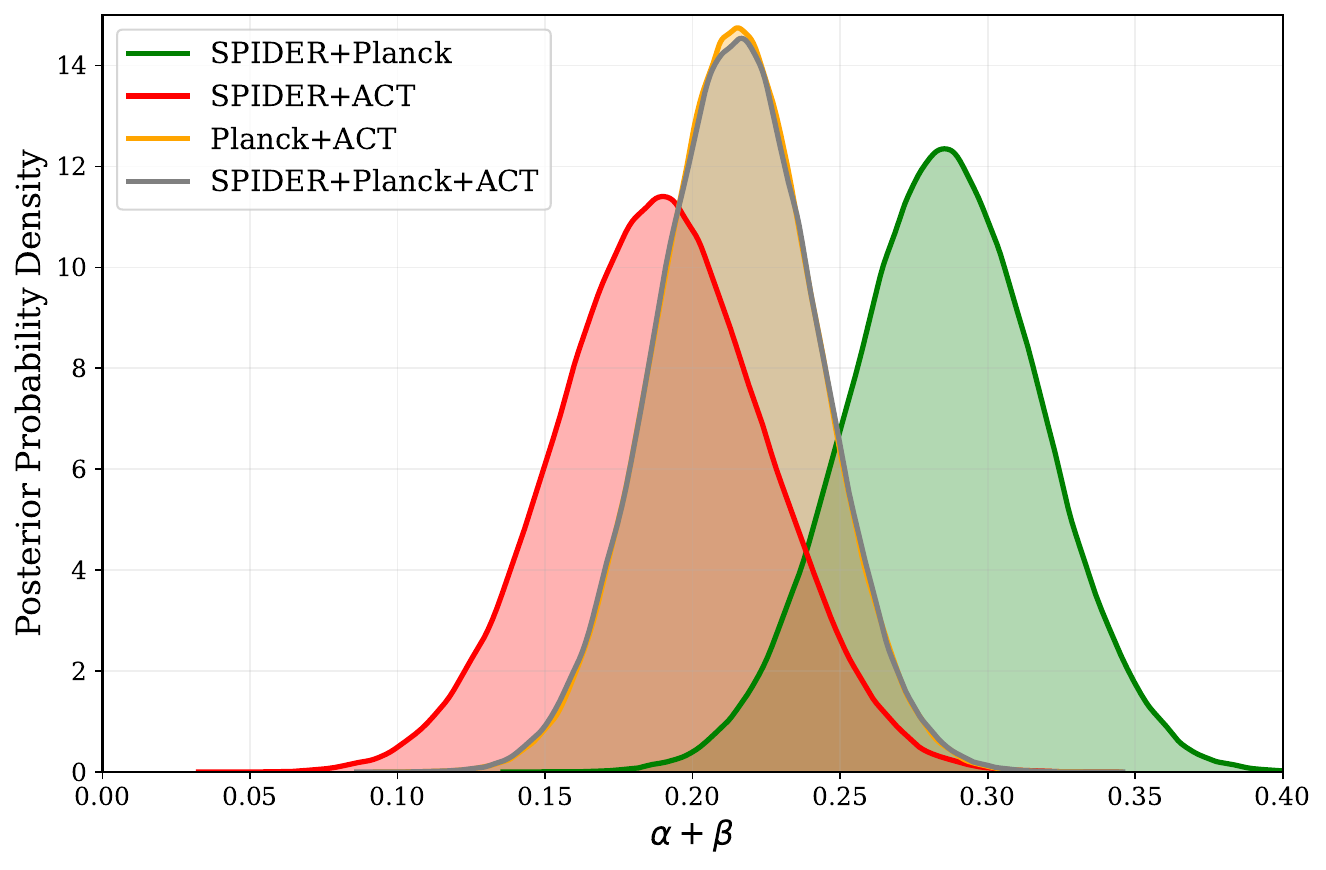}\\
(a)  & (b)   \\
    \end{tabular}
\caption{\label{fig:alpha+beta-PDF} 
 {The probability density functions for a constant rotation angle, $\alpha+\beta$, using the independent CMB $EB$ measurements from SPIDER, Planck, and $EB$ ($TB$) measurements from ACT (a), and the corresponding results using different combinations of the datasets (b).}
}
\end{figure}

In this section, we disregard the specific {EDE} model and focus instead on the model-independent constraint on the total rotation angle $\alpha + \beta$. Here, $\alpha$ represents the instrumental miscalibration angle, mainly caused by imperfections in the {calibration of} individual detectors, while $\beta$ denotes the cosmological birefringence rotation angle. Since $\alpha$ and $\beta$ cannot be fully disentangled at present, the observational constraints are placed on their sum $\alpha + \beta$. {While $\alpha$ can be different for each experiment, $\beta$ is expected to remain the same across all experiments since it is of cosmic origin. Hence, if the measured rotation angles across different experiments are similar, then that provides evidence that the rotation angle may have a stronger contribution from cosmic birefringence rather than polarization miscalibration.}
We find that the SPIDER and Planck data individually constrain $\alpha + \beta$ to be ${0.35^\circ}\pm{0.69^\circ}$ and $0.29^\circ\pm 0.03^\circ$, respectively. 
Interestingly, although the ACT-$EB$ and ACT-$TB$ analyses yield different values for the birefringence coupling parameter $gM_{Pl}$, the central values of $\alpha + \beta$ are consistent between the two, indicating that a constant rotation angle may be the more likely explanation for the observations.

In the combined data analysis, we cut the Planck data to exclude overlaps with SPIDER or ACT to avoid double-counting. We use the lower cut $\ell_{min}$=250 and the upper cut $\ell_{max}$=600 for the Planck data, when combining it with SPIDER and ACT data respectively. We repeated all our analyses without applying any data cuts and found that it does not significantly change the results and conclusions. For consistency, we use the Planck data cuts throughout the article whenever analyzing combinations of datasets. Across all data combinations, we find that $\alpha + \beta$ remains positive, and {we calculated the confidence of} it does not return to zero. {In SPIDER-$EB$, Planck-$EB$, ACT-$EB$, and ACT-$TB$ datasets, the confidence of $\alpha+\beta>0$ is 0.51$\sigma$, 9.67$\sigma$, 4.75$\sigma$, and 1.73$\sigma$, respectively. For the combined dataset, the confidence of $\alpha+\beta>0$ in SPIDER+Planck, SPIDER+ACT, Planck+ACT, SPIDER+Planck+ACT are 9.67$\sigma$, 4.75$\sigma$, 7.0$\sigma$, and 7.3$\sigma$, respectively.} 
The corresponding $\chi^2$, PTE, and posterior probability density are shown in Table \ref{tab:alpha+beta} and Figure \ref{fig:alpha+beta-PDF}. We note that the constraint on $\alpha + \beta$ is almost identical between the Planck+ACT and SPIDER+Planck+ACT data combinations, which suggests that most of the constraining power comes from Planck and ACT, and SPIDER contributes very little towards the constraints.

\begin{figure}
\centering   
\begin{tabular}{cc}
\includegraphics[width=0.5\linewidth]{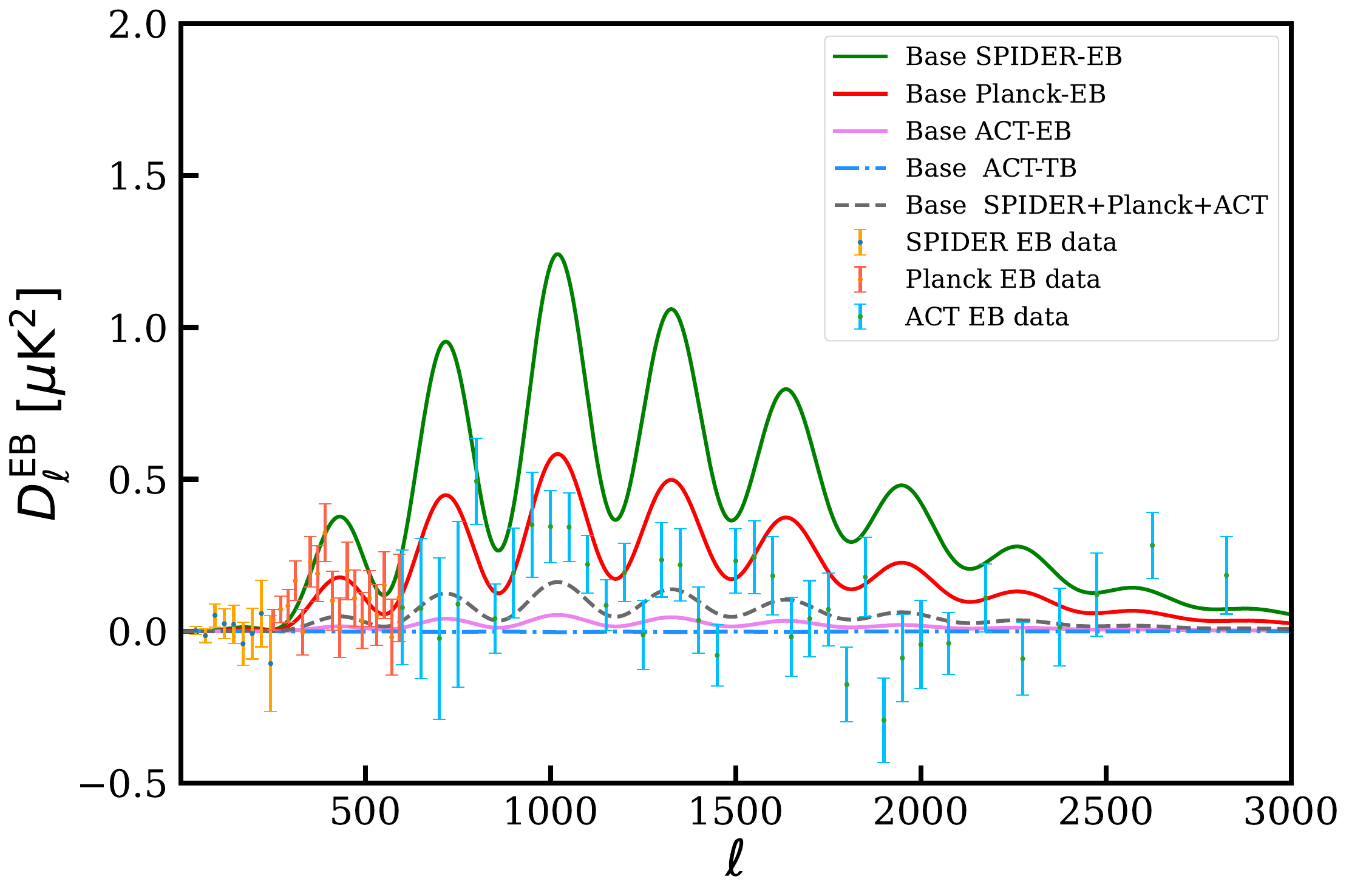}&
\includegraphics[width=0.5\linewidth]{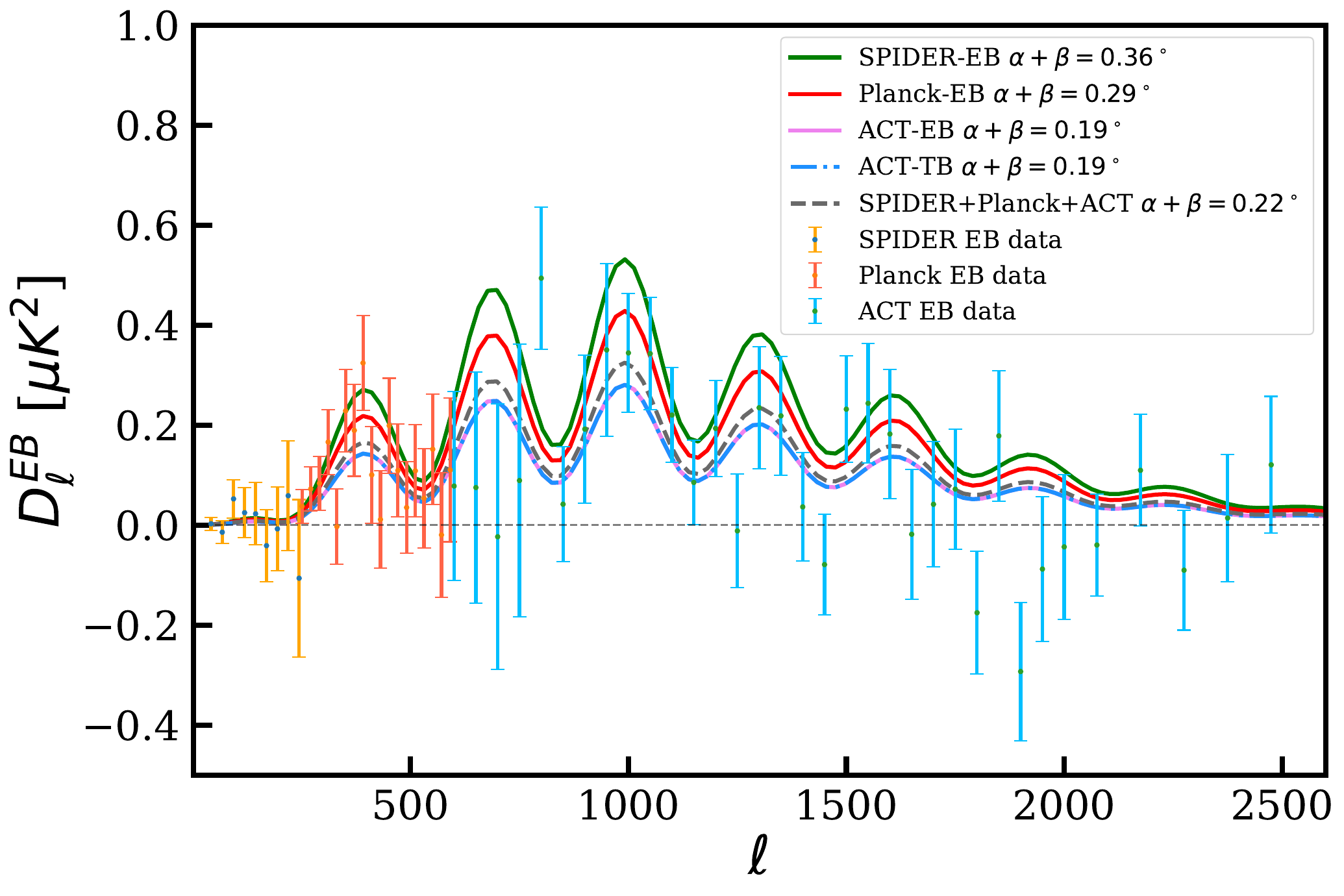}\\
(a)  & (b)   \\
    \end{tabular}
\caption{\label{fig:5} 
{The summary result of CMB-$EB$ power spectra for the individual datasets of SPIDER, Planck, and ACT. The left-hand side (a) shows that during the other EDE parameters in Base, Base+SH0ES, and BSL results, the green, red, pink, blue, and gray lines  
come from the constraints of Chern-Simons coupling constant $gM_{Pl}$ in SPIDER-$EB$, Planck-$EB$, ACT-$EB$, ACT-$TB$, and SPIDRER+Planck+ACT datasets, respectively. The right-hand side (b) shows the model-independent $\alpha+\beta$ results in corresponding $EB$ datasets.  }
}
\end{figure}

Fig. \ref{fig:5} provides a summary and comparative overview of the results above. {The theoretical best-fit curves from SPIDER and Planck data are extrapolated to high multipoles for the purpose of comparison.} Left-hand side figure (a) can be clearly seen that the best-fit values from SPIDER $EB$ at higher $l$ (from 500 to 2000) deviate significantly from the confidence intervals of ACT and Planck data 1 $\sigma$ range, showing that only using SPIDER at low-$l$ data provides weak constraints. We also show combined constraints using all three groups of data, represented by a dashed gray line. The constraint numbers are given in the Table \ref{tab:chi2-2}.
Figure (b) on the right-hand side shows the combined results of the model-independent angle $\alpha+\beta$ in different datasets. The corresponding values are shown in Table \ref{tab:alpha+beta}.

\begin{figure}
\centering
    \begin{tabular}{cc}
\includegraphics[width=0.5\linewidth]{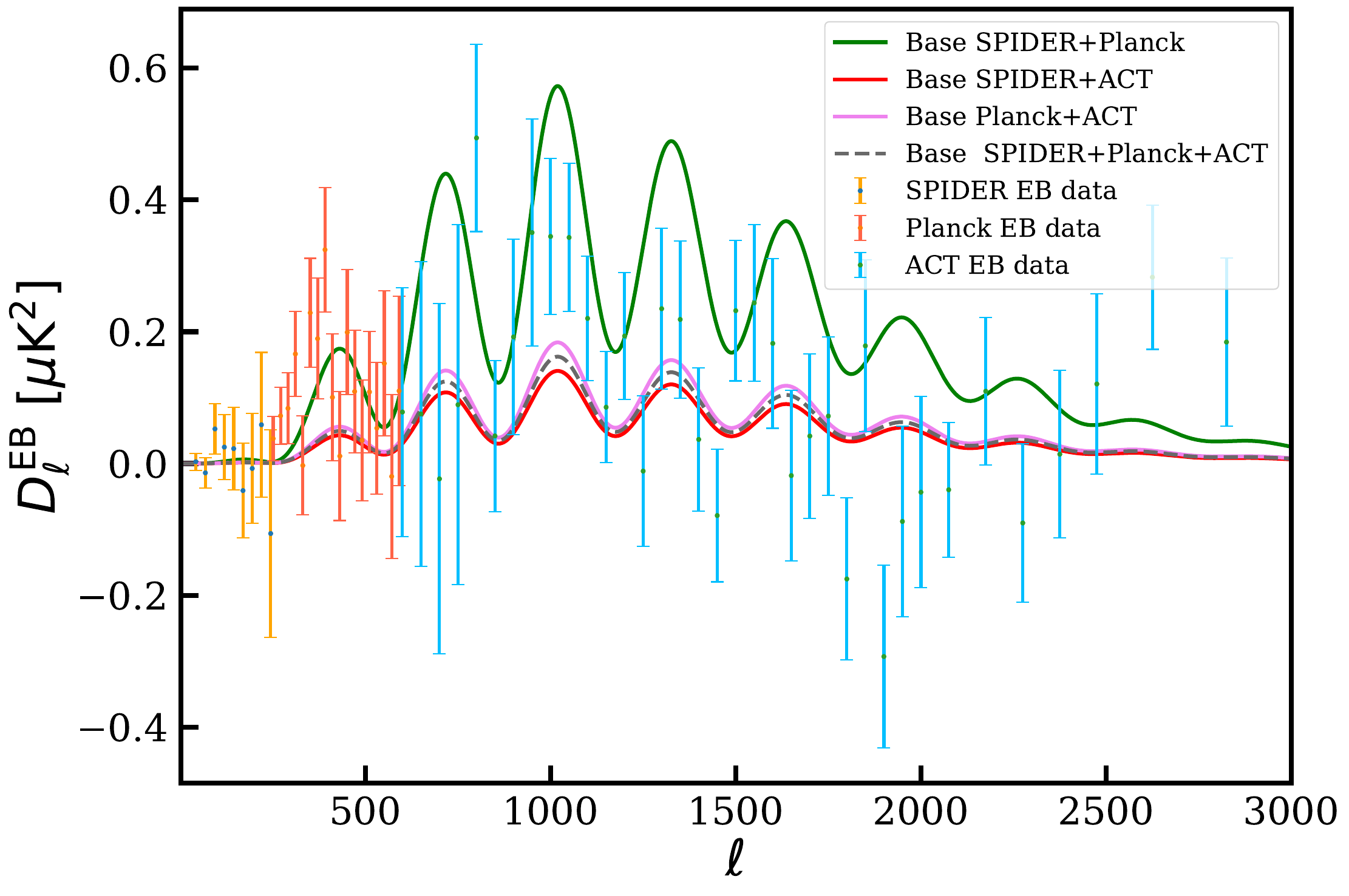}&
\includegraphics[width=0.5\linewidth]{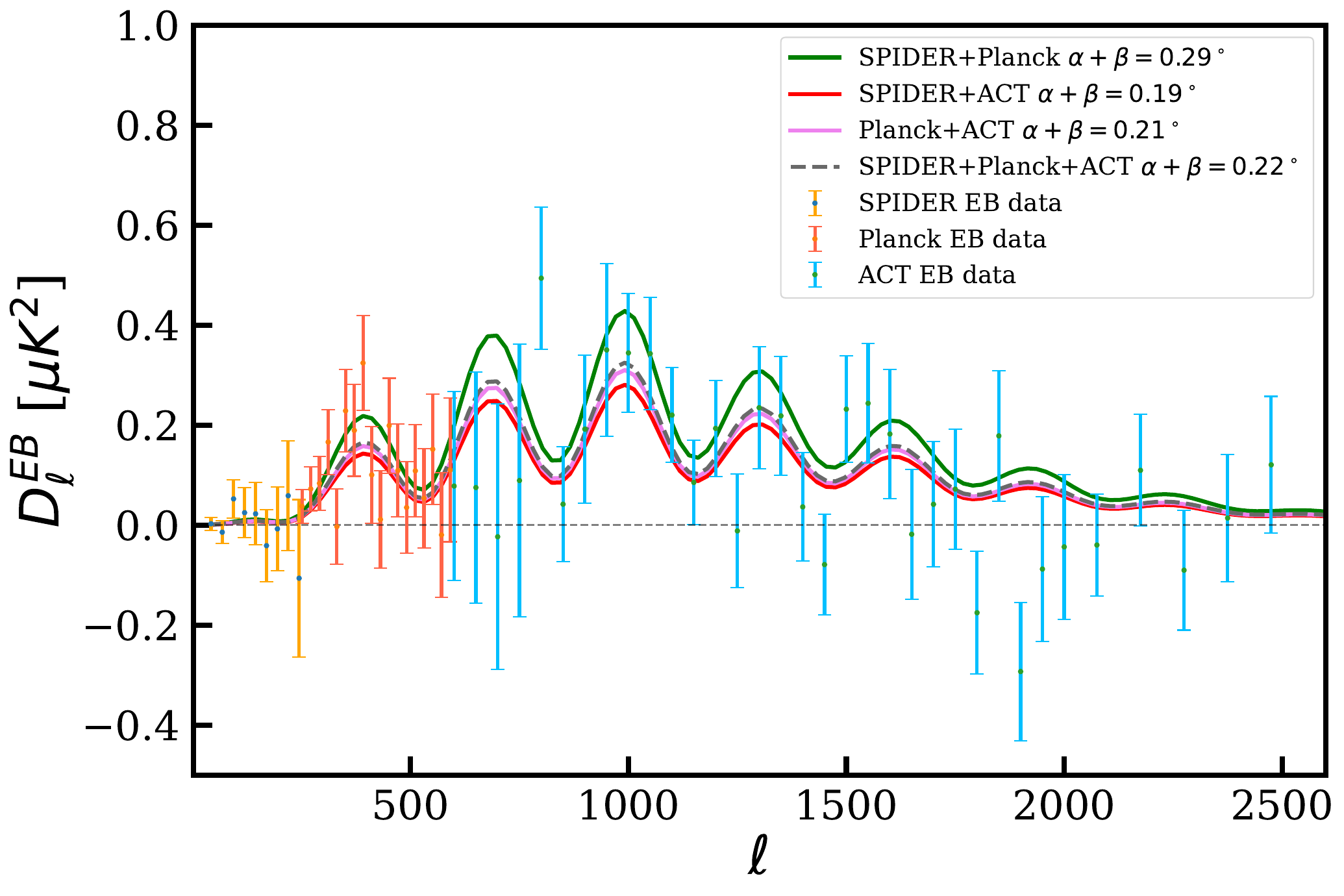}\\
(a)  & (b)   \\
    \end{tabular}
\caption{\label{fig:6} The summaries result of CMB-$EB$ power spectra {for combinations of the SPIDER, Planck and ACT datasets}. The left-hand side (a) shows that during the other EDE parameters in Base, Base+SH0ES, and BSL results, the green, red, pink, and gray lines  
come from the constraints of Chern-Simons coupling constant $gM_{Pl}$ in SPIDER+Planck, SPIDER+ACT, Planck+ACT, and SPIDRER+Planck+ACT datasets, respectively. The right-hand side (b) shows the model-independent $\alpha+\beta$ results in the corresponding $EB$ datasets. }
\end{figure}

To evaluate whether pairwise combinations of datasets improve the constraints, we also tested several combinations: SPIDER+Planck, SPIDER+ACT, and Planck+ACT. The resulting $gM_{Pl}$, $\chi^2$, and PTE values are shown in Table \ref{tab:chi2-2} and Fig. \ref{fig:6}(a). For combined SPIDER+Planck+ACT, under Base dataset, $gM_{pl} = 0.15$ with $\chi^2 = 191.00$; under Base+SH0ES, $gM_{pl} = 0.25$ with $\chi^2 = 188.41$. And under BSL, $gM_{pl}=0.02$ and $\chi^2=218.98$. The corresponding PTE values are smaller than $10^{-13}$, showing that $gM_{Pl}$ does not provide a good enough fit to these three data combinations.  Future studies will require both higher-precision measurements and a greater volume of low-multipole (low-$l$) data to achieve more robust constraints. Comparing $gM_{Pl}$ and PTE result by only using ACT-$EB$ data in Table \ref{tab:chi2-single} with Planck+ACT result in Table \ref{tab:chi2-2}, we can find the fitting result by Planck+ACT has higher confidence than use ACT-$EB$ only, that shows high-$l$ observation results need to combine with Planck to get better analyze in the cosmic birefringence. 
The result of $\alpha+\beta$ are shown in Table \ref{tab:alpha+beta} and Fig. \ref{fig:6} (b). Notably, the Planck+ACT combination dataset PTE for $\alpha+\beta$ is 0.490, and  PTE for $gM_{Pl}$ is $1.72\times10^{-3}$, $2.20\times10^{-4}$, and $0.0532$ in Base, Base+SH0ES, and BSL datasets, respectively. These results are significantly more robust than other combined datasets, demonstrating that the Planck+ACT combination will be a more reliable dataset in the future. 
The results also indicate that the ACT-$EB$ data set produces a higher PTE when combined with Planck. In some cases, it could even exceed the confidence level achieved by using ACT data alone.

\begin{table}[]
    \centering
\begin{tabular}{|c|c|c|c|c|}
\hline & & Base & Base+SH0ES & { BSL} \\
\hline 
\multirow{2}{*}{SPIDER+Planck} &{$gM_{Pl}$} & $0.53$ & $0.62$ &  $0.15$\\
  \cline{2-5}
   & $\chi^2$ & $78.36$ & $100.77$ & $67.59$\\
\cline{2-5}
                     & $\nu$      & $  {70}$ &$ 70$ & { $ 70$}\\
\cline{2-5}
     & PTE  & $  0.2308$ &$  0.009396$ & $  0.5596$\\
\hline 
\hline 
\multirow{2}{*}{SPIDER+ACT} & {$gM_{Pl}$} & $0.13$ & $0.14$ & $0.06$\\
    \cline{2-5}
    & $\chi^2$ & $79.62$ &$84.62$ & { $68.35$}\\
\cline{2-5}
                     & $\nu$      & $ 46$ &$ 46$ & { $ 46$}\\
  \cline{2-5}
\cline{2-5}
     & PTE    & $0.001531$ &$4.525\times10^{-4}$ & $ 0.01785$\\
\hline 
\hline 
\multirow{2}{*}{Planck+ACT} & {$gM_{Pl}$} & $0.17$ & $0.19$ & { $0.08$}\\
    \cline{2-5}
    & $\chi^2$ & $103.42$ & $112.79$ & { $84.41$}\\
\cline{2-5}
                     & $\nu$      & $  {65}$ &$ 65$ & { $ 65$}\\
\cline{2-5}
     & PTE    & $  {1.72\times10^{-3}}$ &$   {2.20\times10^{-4}} $ & $ {0.0532}$\\
\hline
\hline 
\multirow{2}{*}{SPIDER+Planck+ACT} & {$gM_{Pl}$} & $0.15$ & $0.25$ & $0.02$\\
  \cline{2-5}
  & $\chi^2$ & $191.00$ & $188.41$ & $218.98$\\
\cline{2-5}
                     & $\nu$      & $ 64$ &$ 64$ & { $ 64$}\\
  \cline{2-5}
\cline{2-5}
     & PTE    & $ 1.436\times10^{-14} $ & $ 3.457\times10^{-14}$ & $ 7.877\times10^{-19}$\\
\hline
\end{tabular}
    \caption{Best fitting results for the Chern-Simons coupling constant $gM_{Pl}$, $\chi^2$, and PTE from the EB dataset of combined SPIDER+Planck, SPIDER+ACT, Planck+ACT, and SPIDER+Planck+ACT, respectively.
}
    \label{tab:chi2-2}
\end{table}

\section{Summary}
\label{sec:4}

In this paper, we analyze the Cosmic Birefringence from the Early Dark Energy interacting with photons in Chern-Simons coupling then
constraints on the Chern-Simons coupling constant $gM_{Pl}$ and the model-independent angle $\alpha+\beta$ by using the SPIDER, Planck, and ACT observations. We systematically evaluate their fitting performance through both $\chi^2$ and PTE statistics.
For individual datasets, Planck exhibits a larger PTE and higher confidence levels, both in measuring $gM_{Pl}$ and $\alpha+\beta$. Regarding $\alpha+\beta$, except for SPIDER and ACT-$TB$ which currently have excessively large uncertainties, all other datasets and combined results indicate that $\alpha+\beta$ is greater than zero within 4.75$\sigma$ confidence level. 
Although ACT-$EB$ and ACT-$TB$ show significantly different results for $gM_{Pl}$, their central values for $\alpha+\beta$ are identical. {Additionally, the rotation angle inferred independently from Planck and ACT differ by $>3 \sigma$,  which hints that a polarisation miscalibration angle may be preferred over early dark energy, by the ACT data}. Among the pairwise combined datasets, the Planck+ACT($EB$) combination demonstrates significantly higher PTE and Posterior Probability Density, making it a more promising dataset combination for future exploration. {Including the SPIDER dataset does not significantly impact the constraints, indicating that the contribution of SPIDER to the constraining power is minor.} 
{We find that for the ACT experiment, which primarily probes the high-$\ell$ region, it is necessary to combine it with the relatively lower-$\ell$ measurements from Planck in order to obtain results with higher confidence. This joint analysis yields a better PTE than using ACT-$EB$ data alone.}

We anticipate that future observations with enhanced precision, such as those from LiteBIRD and AliCPT, will provide more robust and conclusive results, particularly when focusing on their combined results with Planck and ACT, which may yield higher confidence levels.

\section*{Acknowledgements}
We sincerely thank Eiichiro Komatsu and Johannes R. Eskilt for their very helpful discussions.
L.Yin was supported by the Natural Science Foundation of Shanghai 24ZR1424600. J. K. was supported by the MCNS faculty development fund of Manipal Academy of Higher Education. B.-H.L. is partially supported by the National Research Foundation of Korea (NRF) grant RS-2020-NR049598, and Overseas Visiting Fellow Program of Shanghai University.

\section*{Appendix}

\begin{figure}
    \centering
    \begin{tabular}{ccc}
\includegraphics[width=0.3\linewidth]{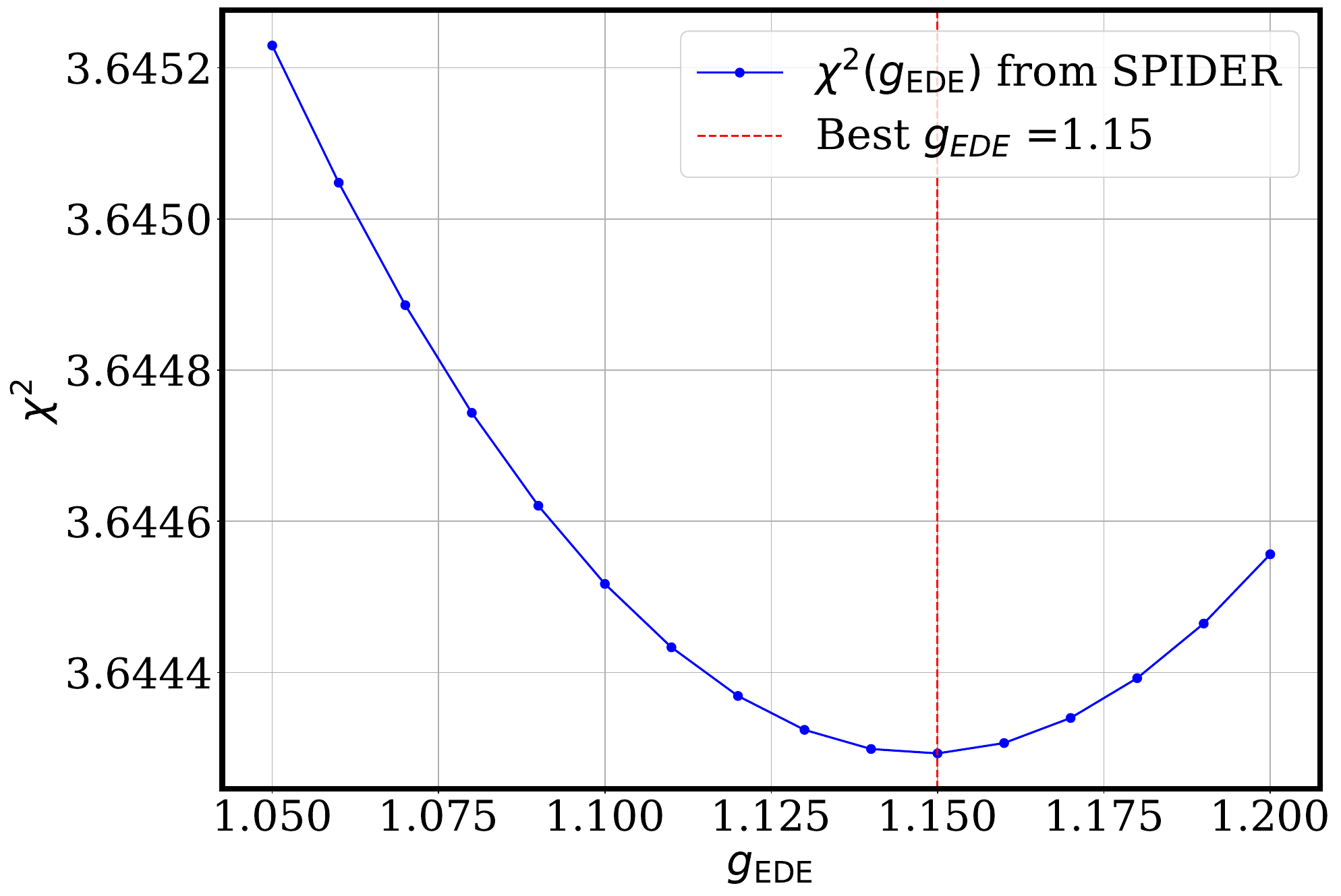}&
\includegraphics[width=0.302\linewidth]{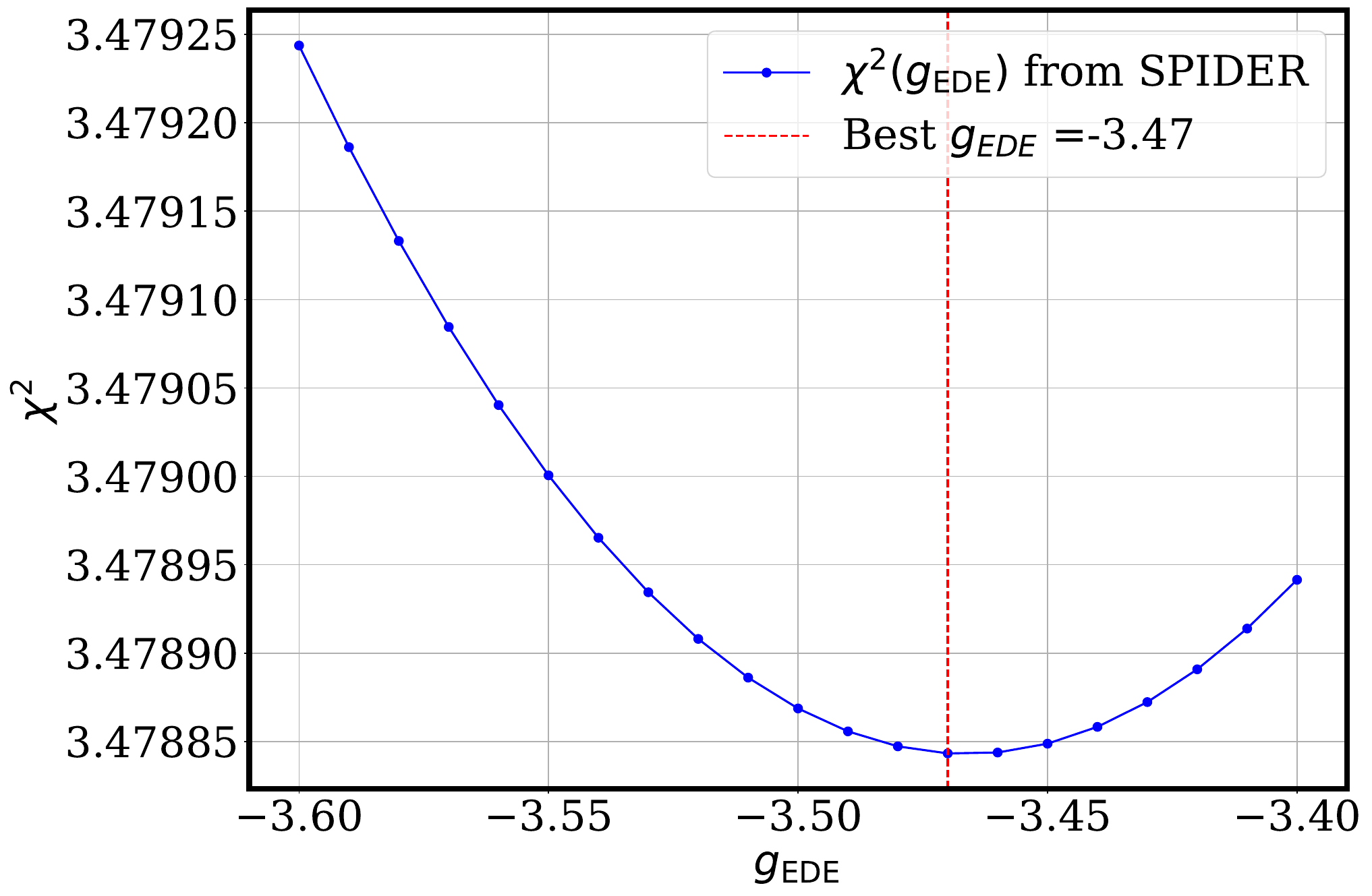}
&
\includegraphics[width=0.295\linewidth]{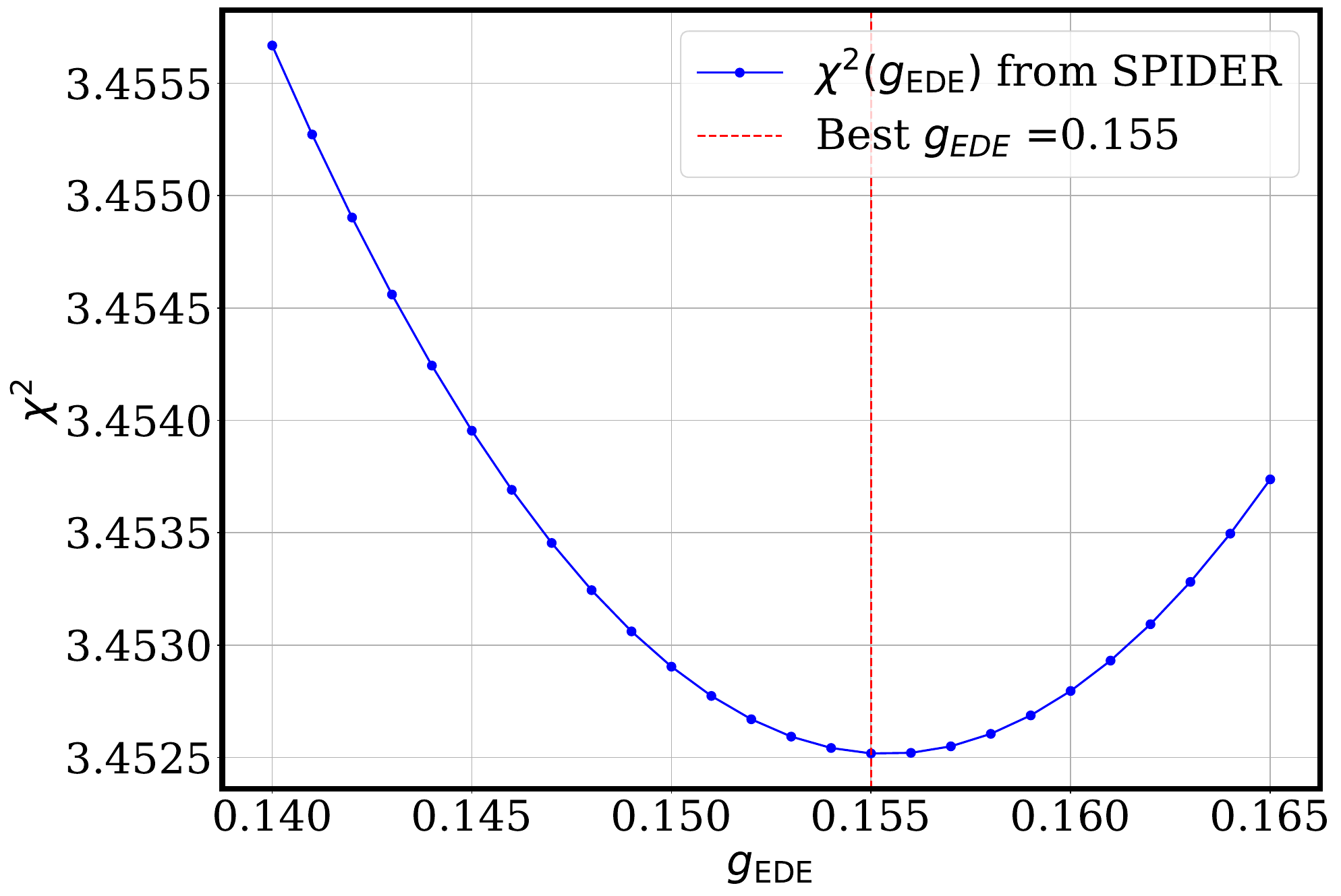}\\
(a)  & (b)  & (c)  \\

\includegraphics[width=0.296\linewidth]{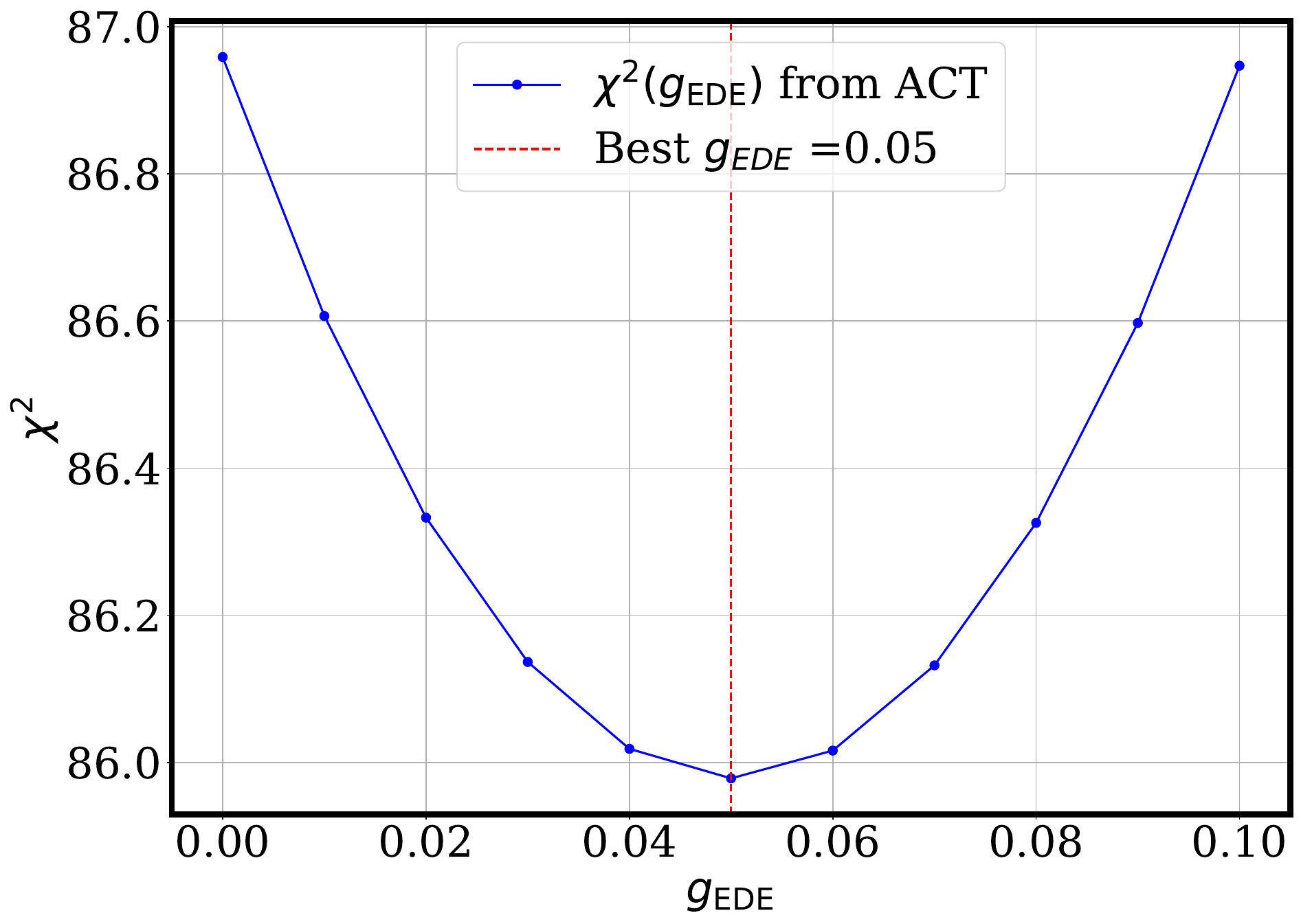}&
\includegraphics[width=0.302\linewidth]{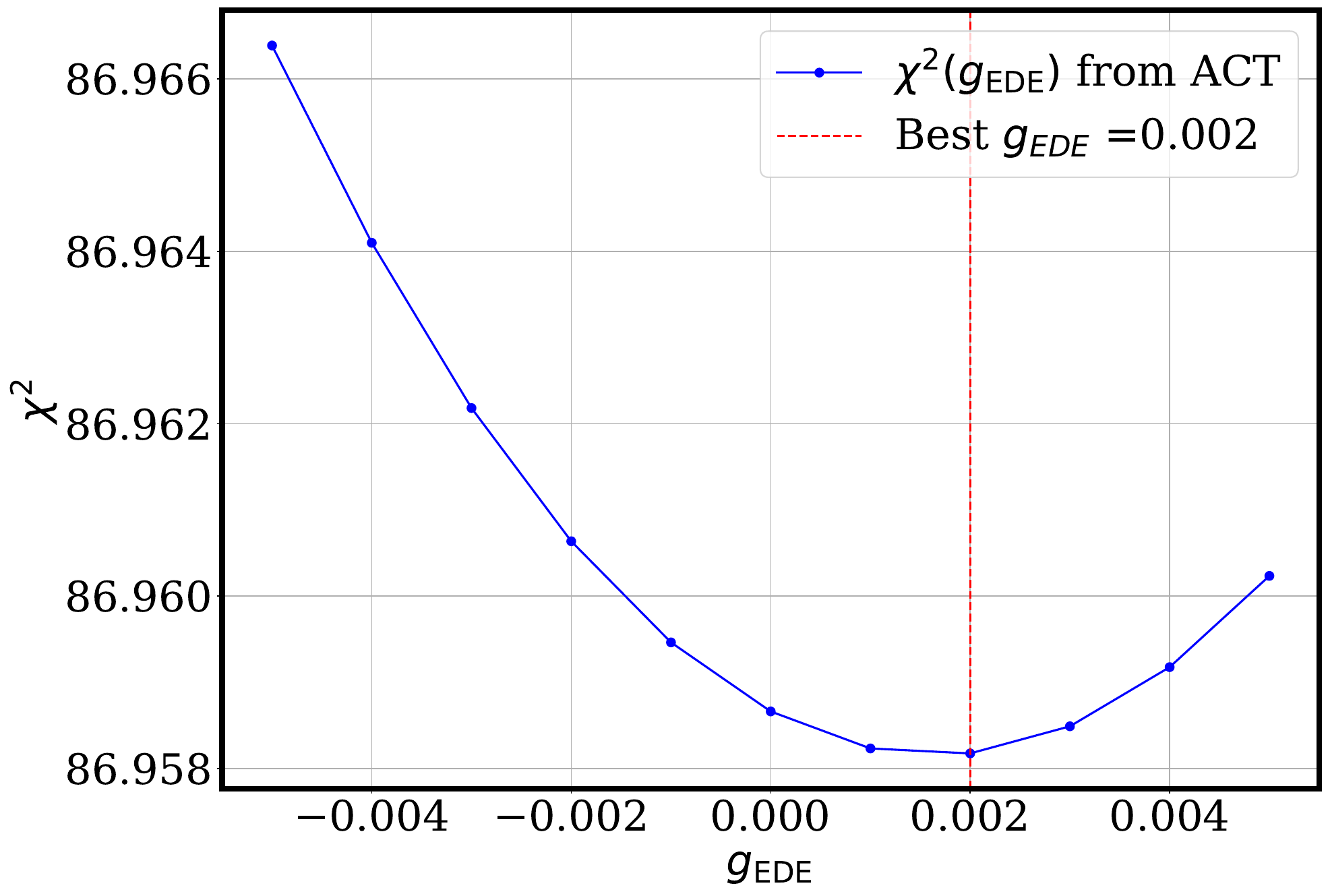}&
\includegraphics[width=0.298\linewidth]{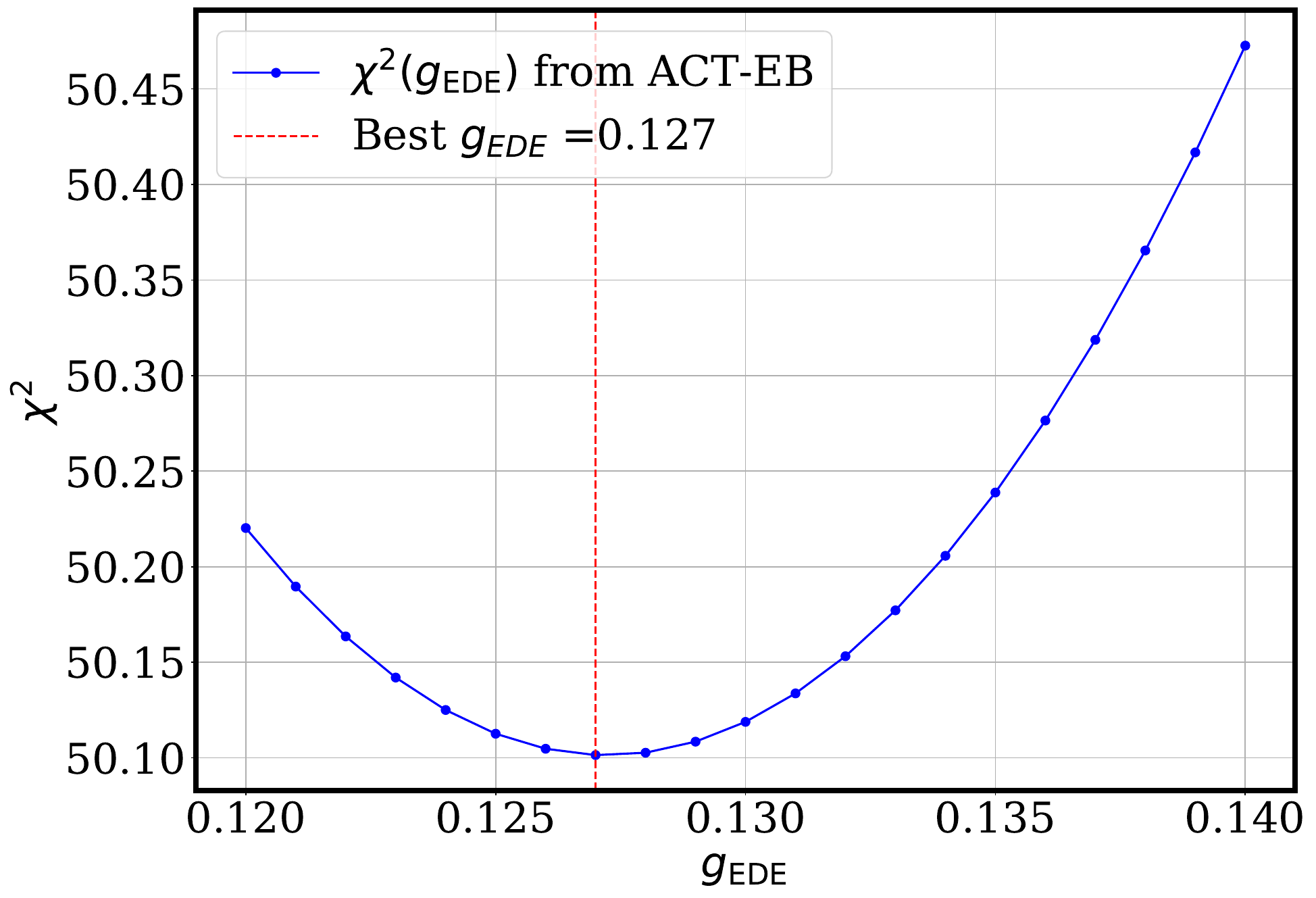}\\
(d)   & (e)  & (f)  \\ 

\includegraphics[width=0.298\linewidth]{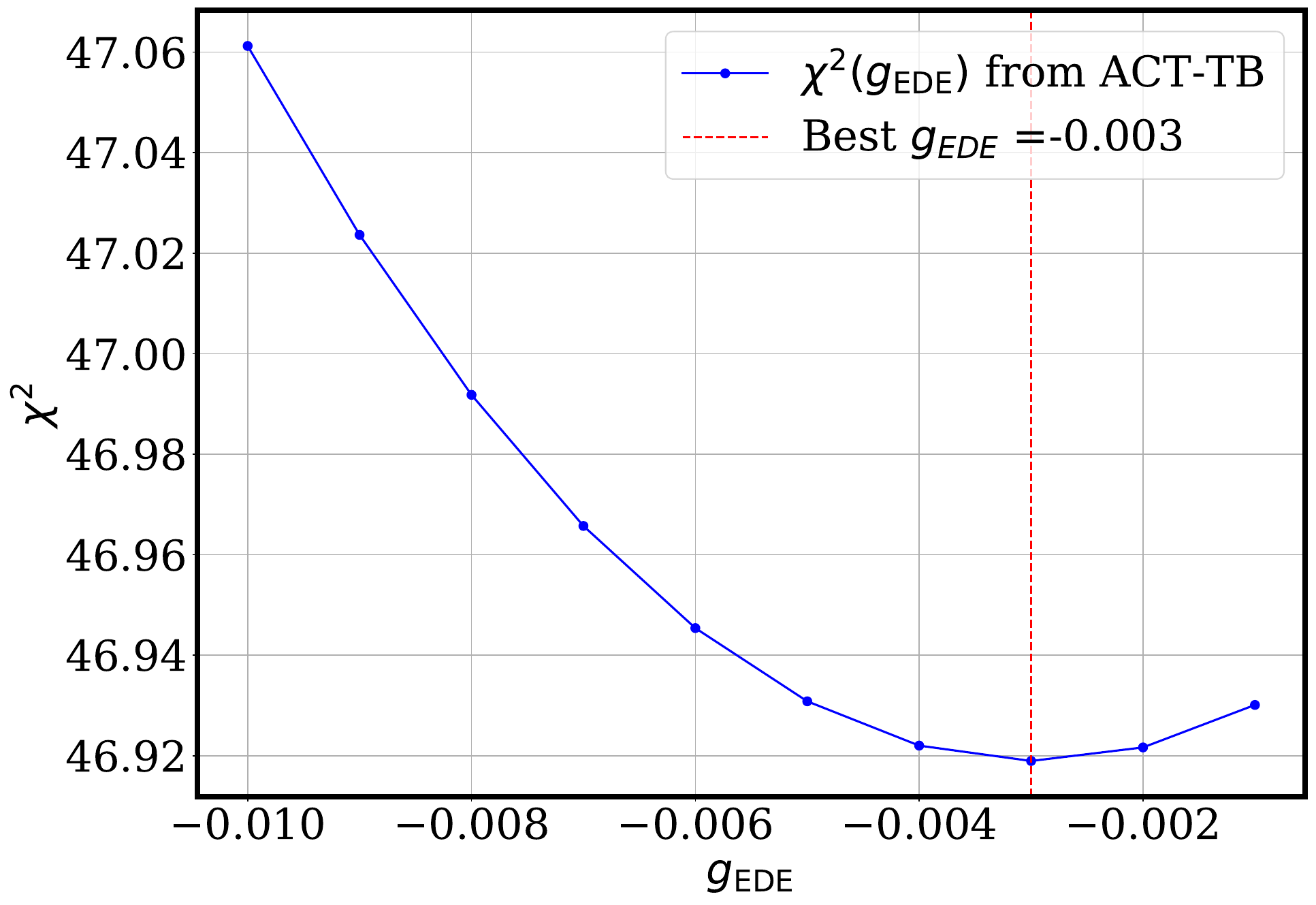}&
\includegraphics[width=0.304\linewidth]{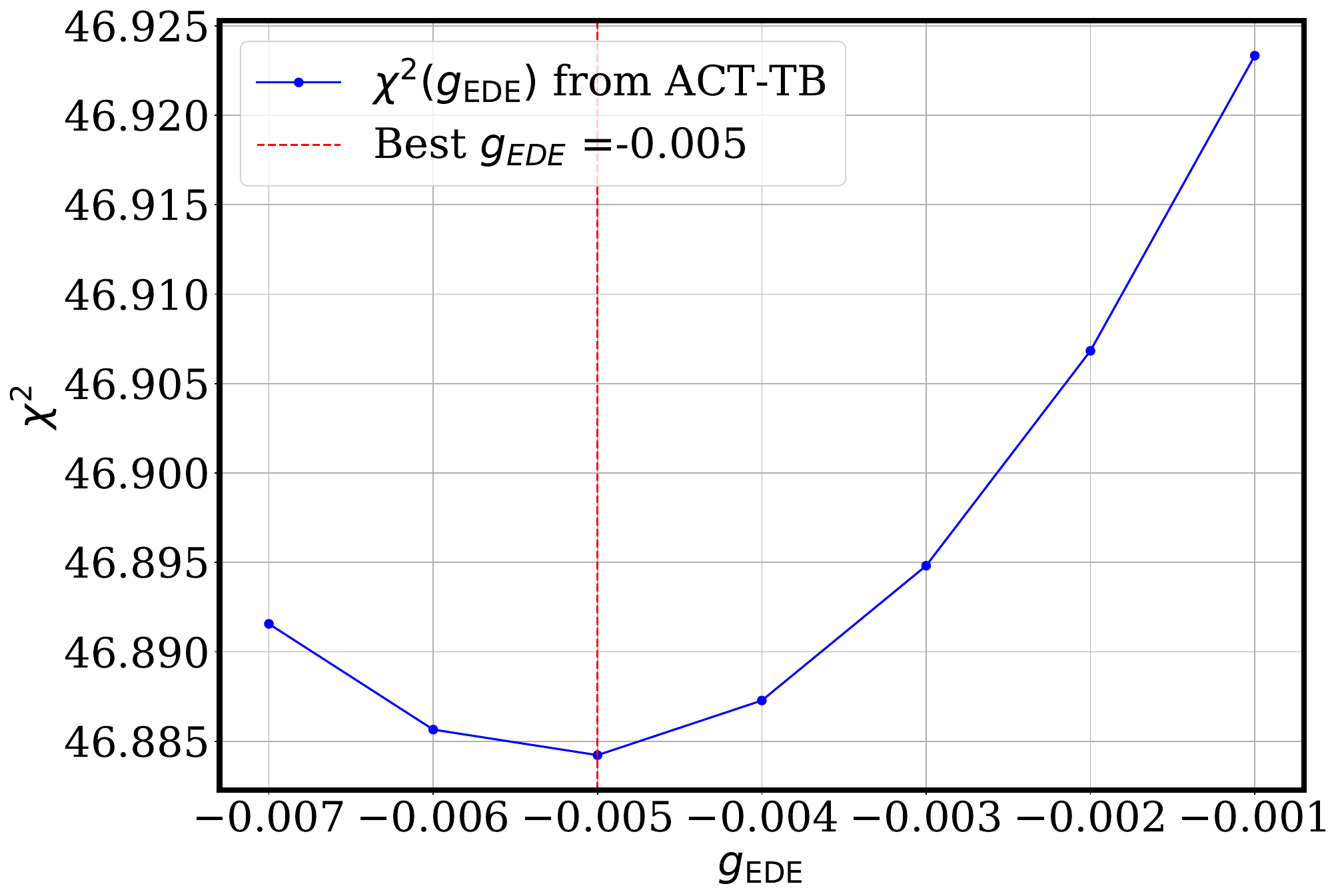}&
\includegraphics[width=0.3\linewidth]{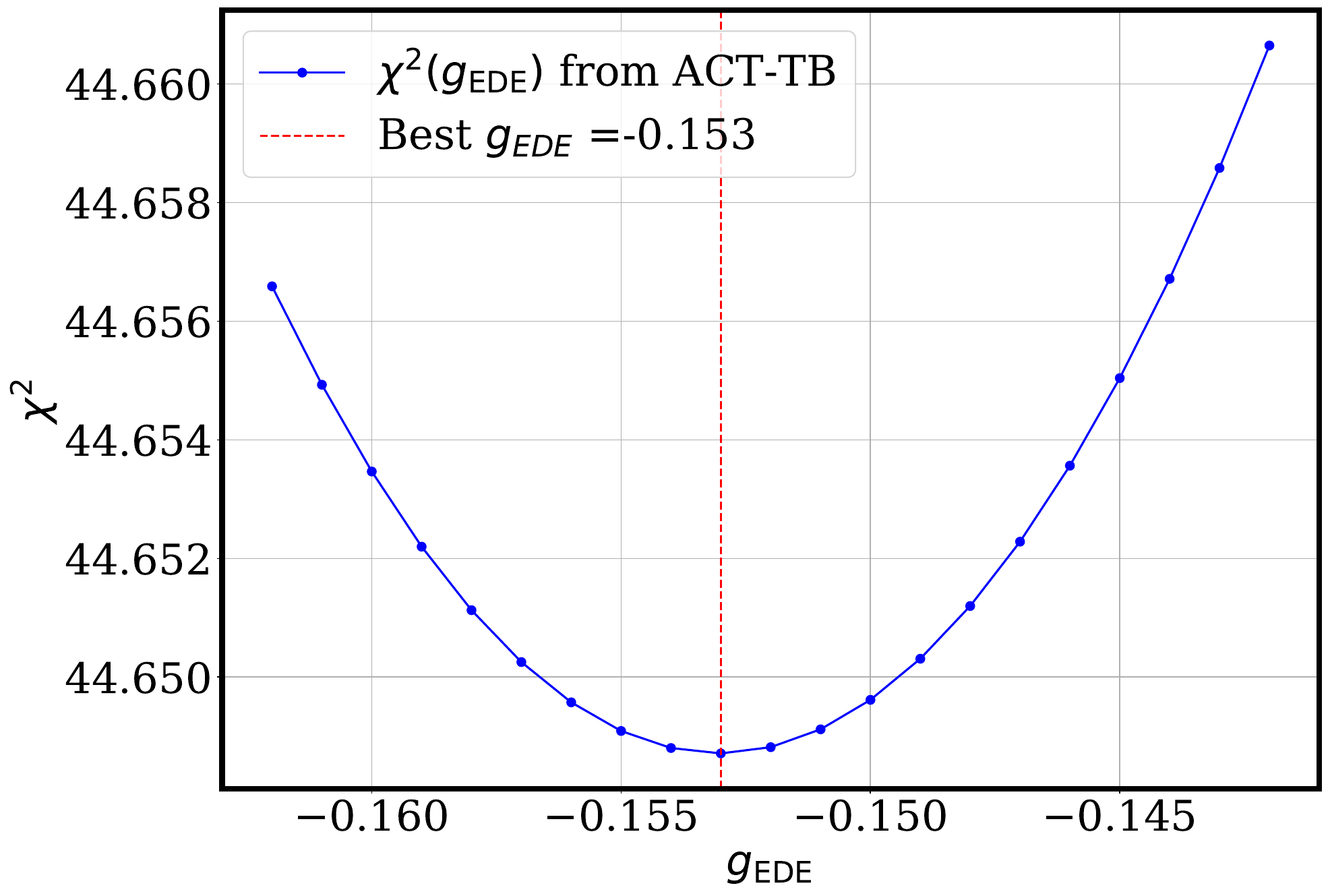}\\
(g)   & (h)   & (i) \\ 

\includegraphics[width=0.3\linewidth]{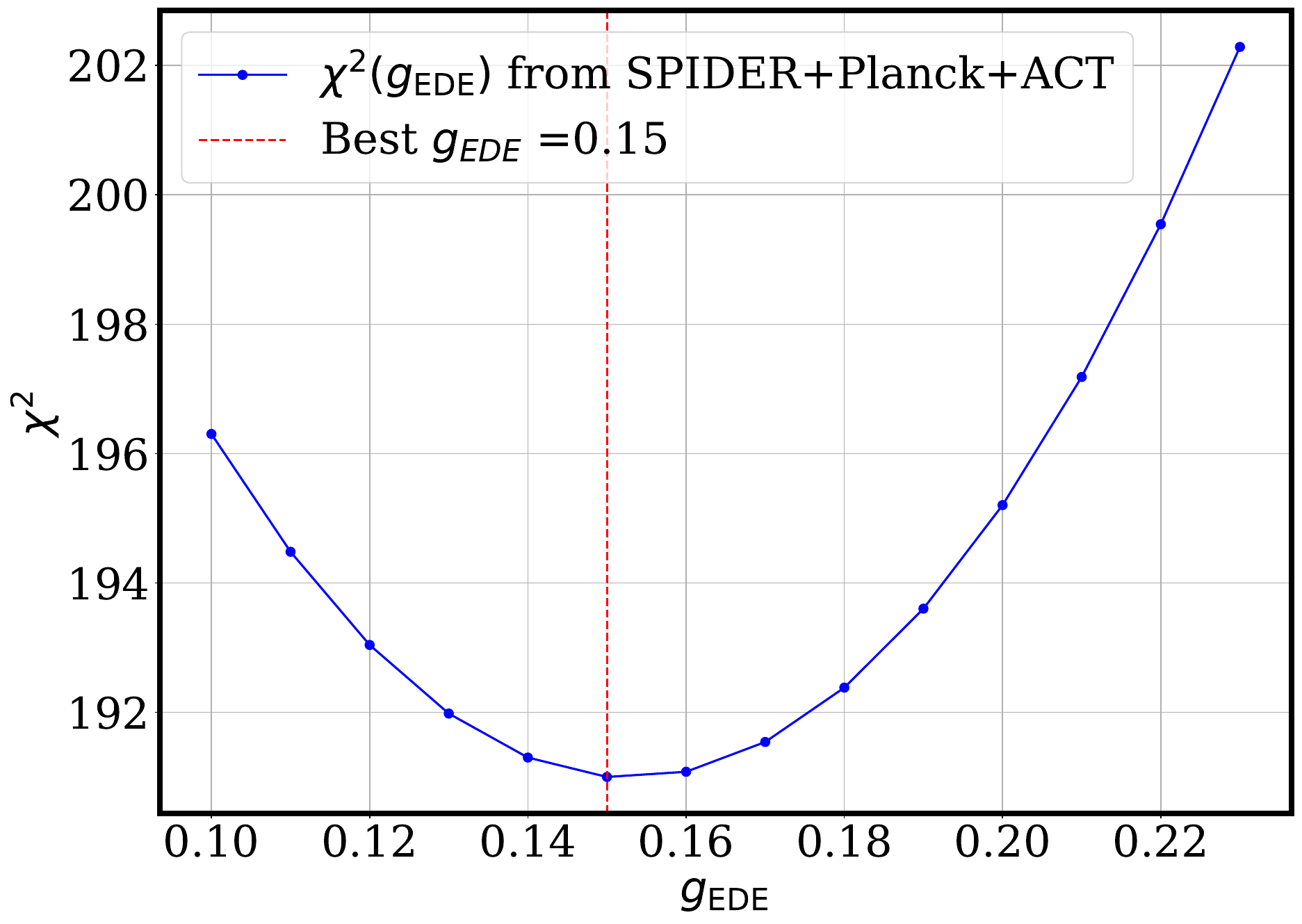}&
\includegraphics[width=0.299\linewidth]{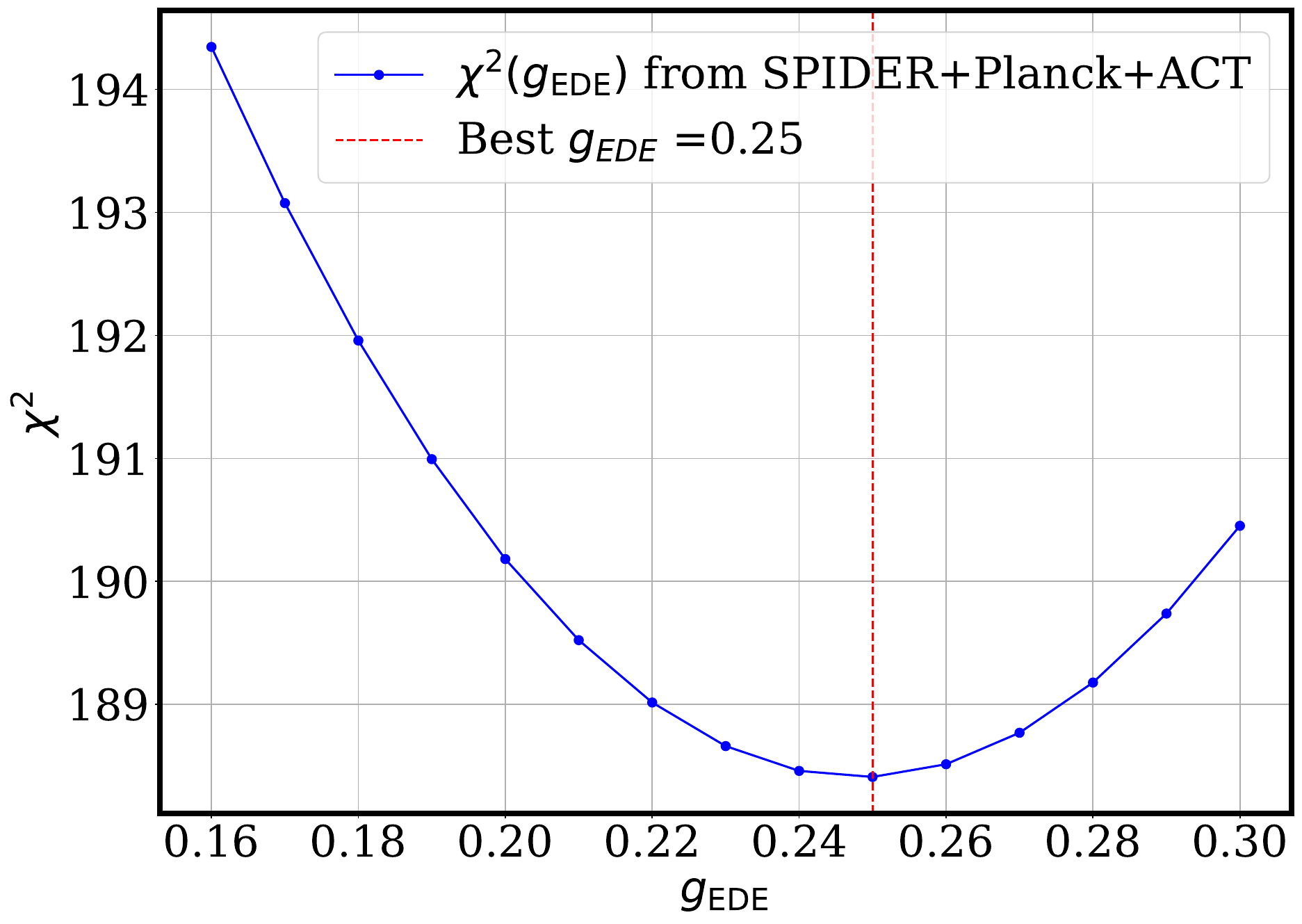}&
\includegraphics[width=0.298\linewidth]{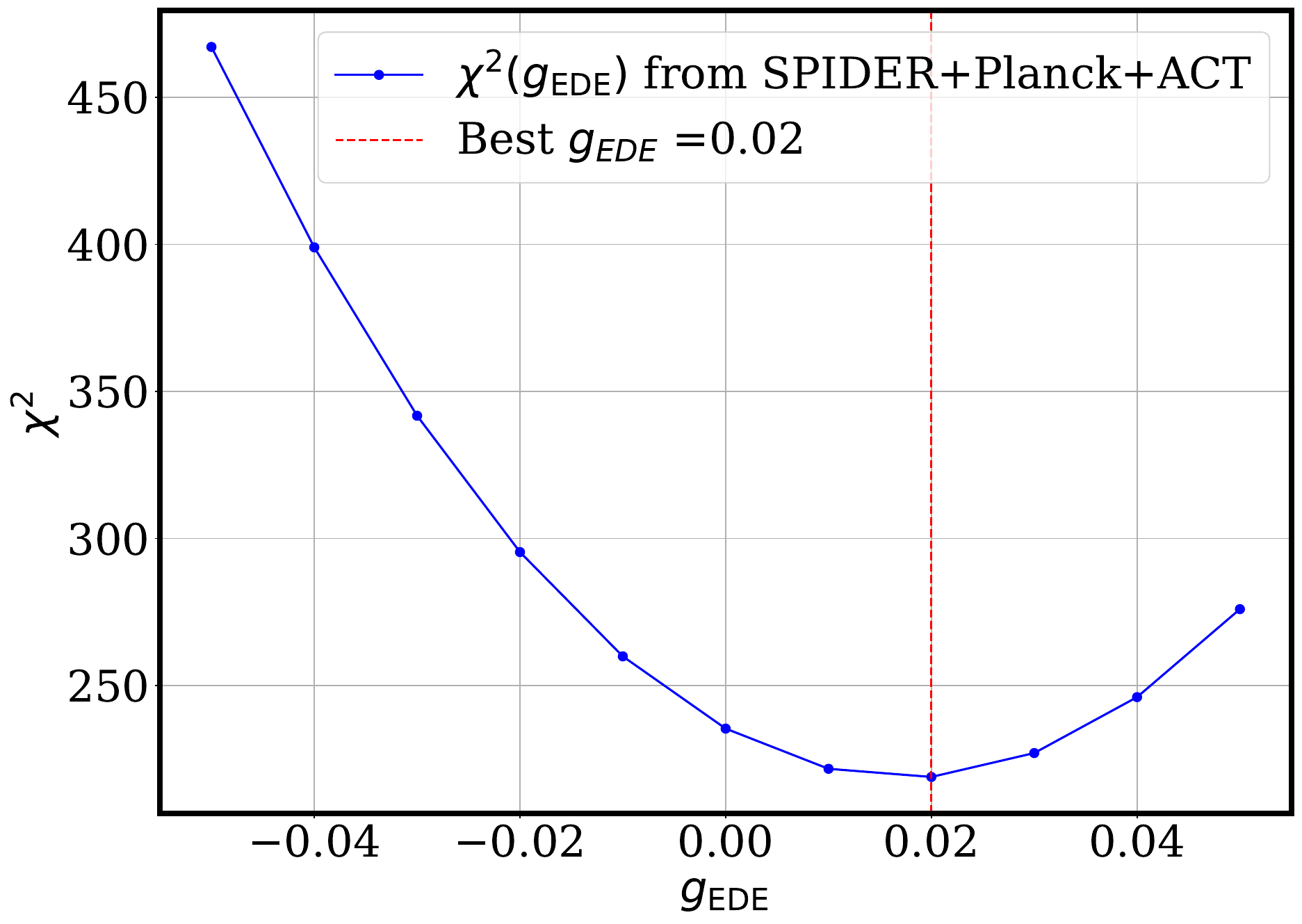}\\
(j)   & (k)  & (l)  \\
    \end{tabular}
\caption{\label{fig:11}{The $\chi^2$ and best-fit of $g_{EDE}$ (the coupling constant $gM_{Pl}$ in EDE model) results from SPIDER, ACT-EB, ACT-TB, and SPIDER+Planck+ACT dataset, respectively. The first, second, and third columns represent the nine fundamental parameters derived from Base, Base+SH0ES, and BSL, respectively. }}
\end{figure}

\begin{figure}[ht]
    \centering
    \begin{tabular}{ccc}
        \includegraphics[width=0.3\linewidth]{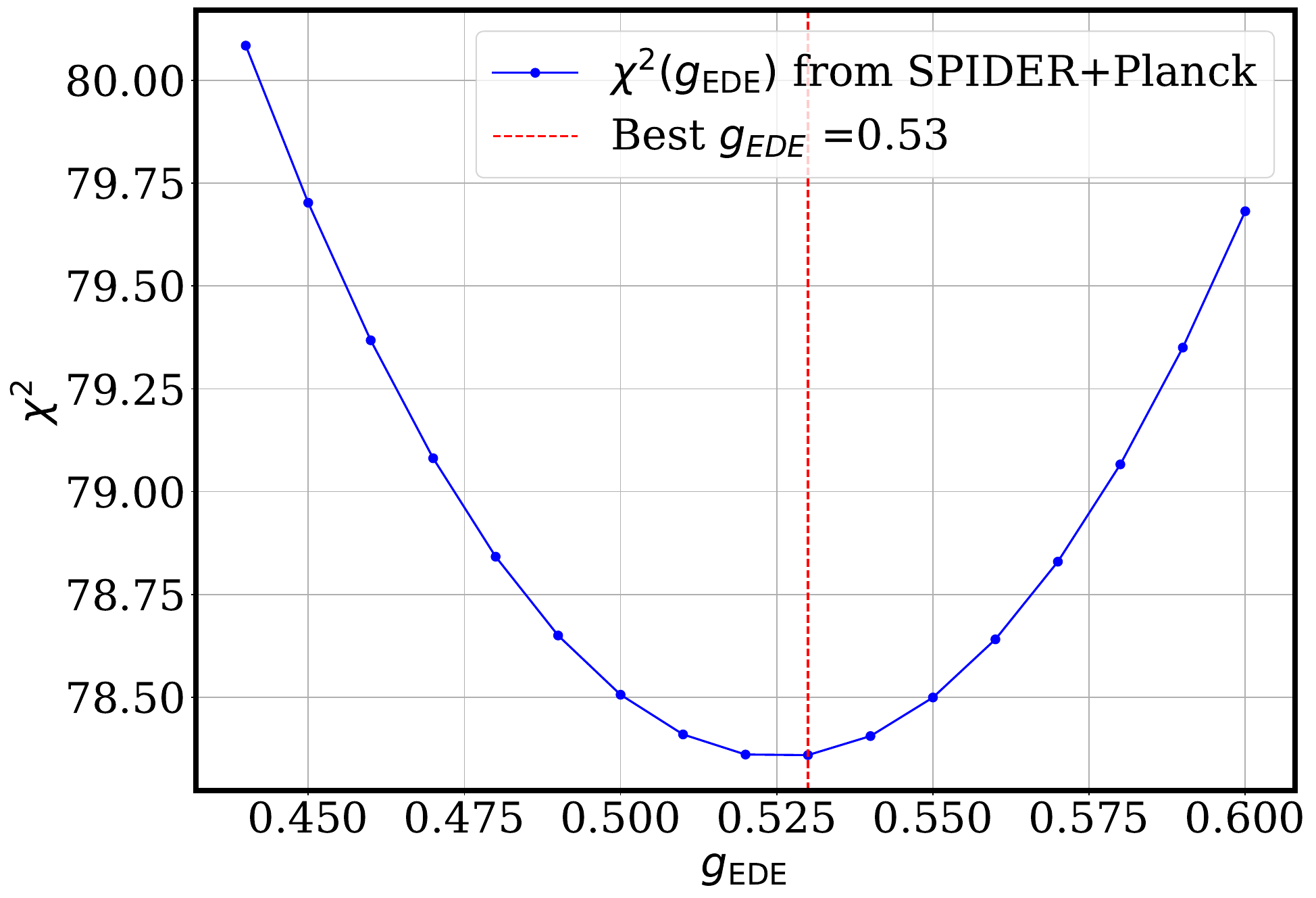} &
        \includegraphics[width=0.3\linewidth]{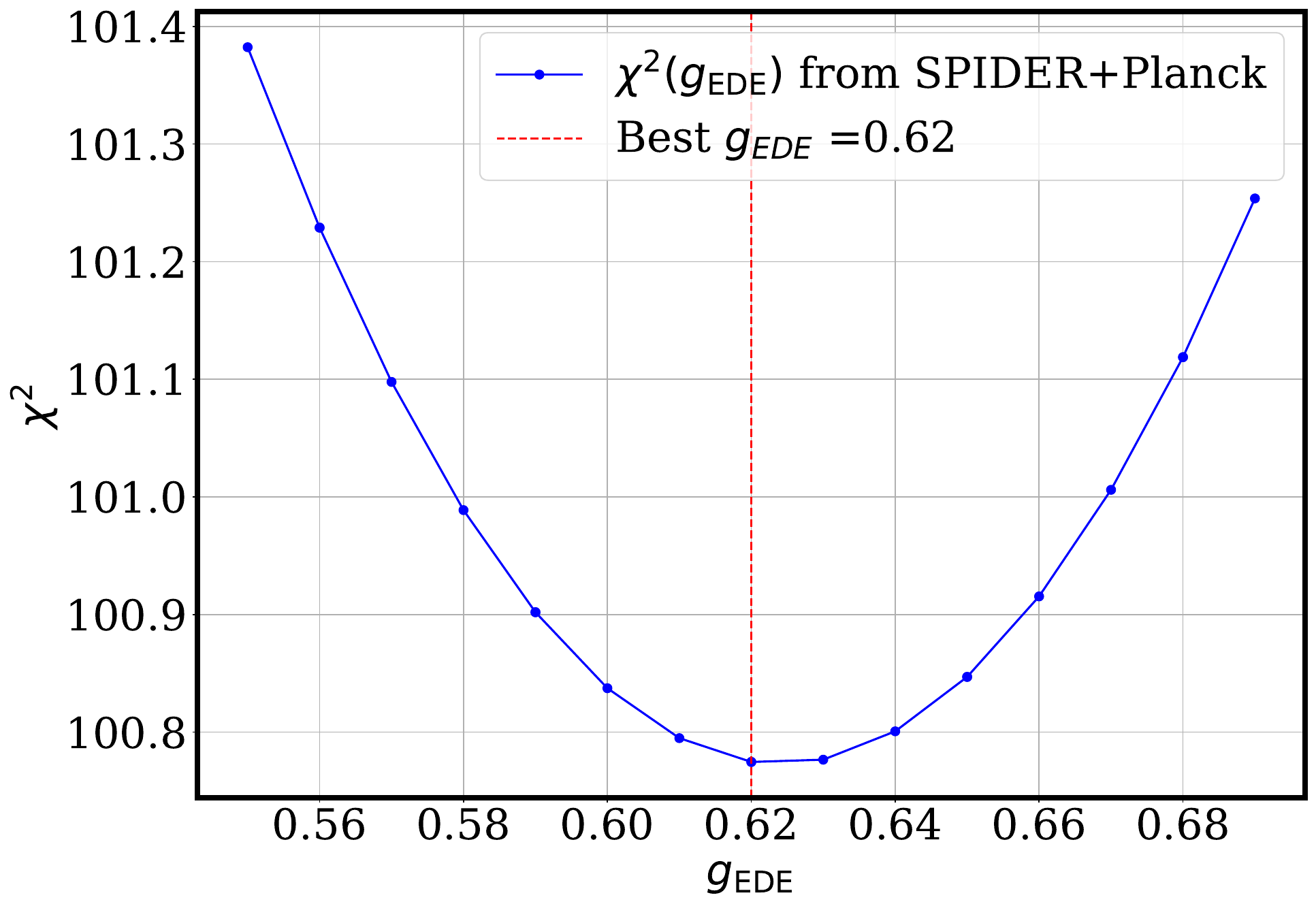}
         &
        \includegraphics[width=0.3\linewidth]{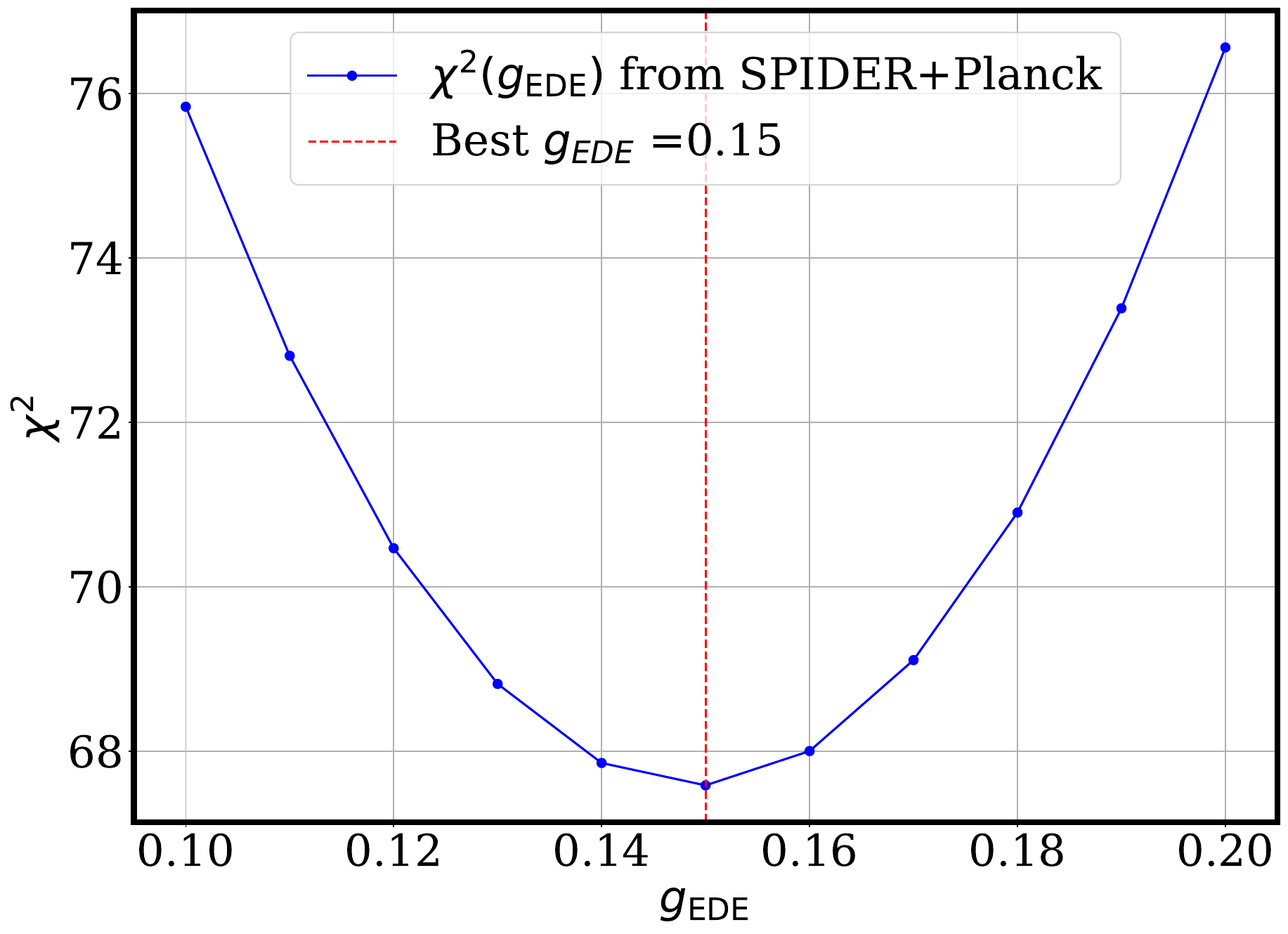} \\
        (a)   & (b)   & (c)  \\ 

        \includegraphics[width=0.3\linewidth]{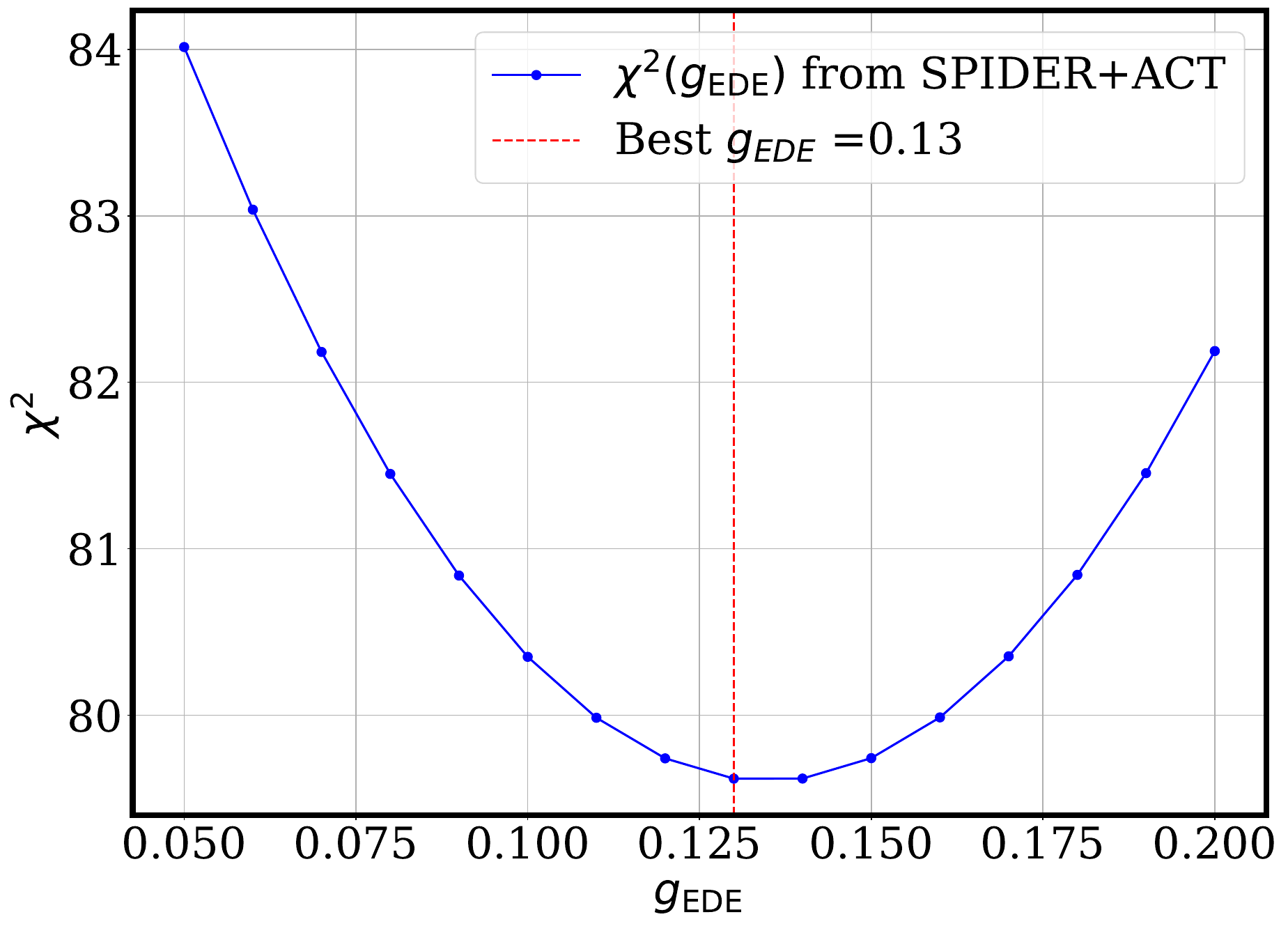} &
        \includegraphics[width=0.31\linewidth]{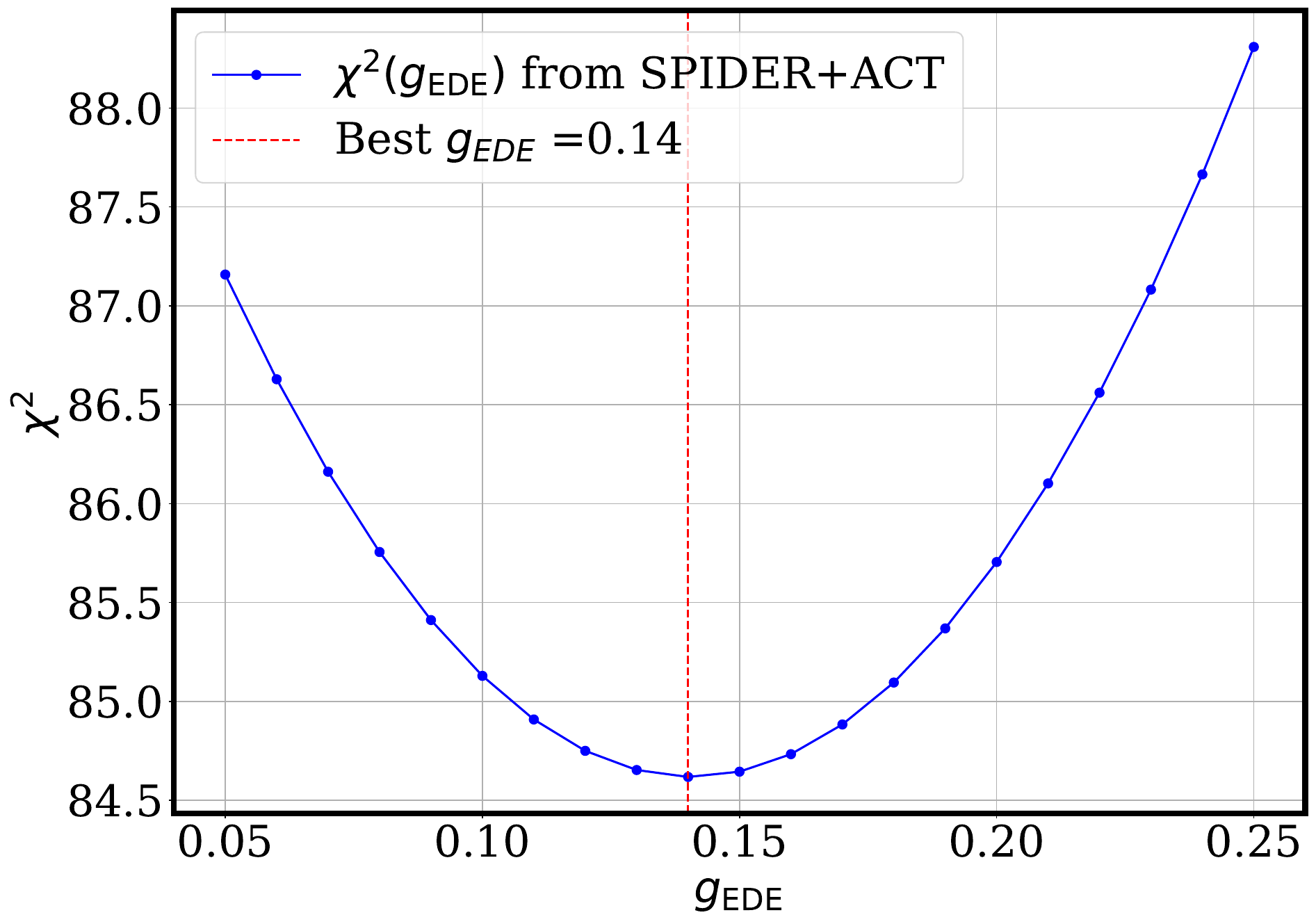} &
        \includegraphics[width=0.3\linewidth]{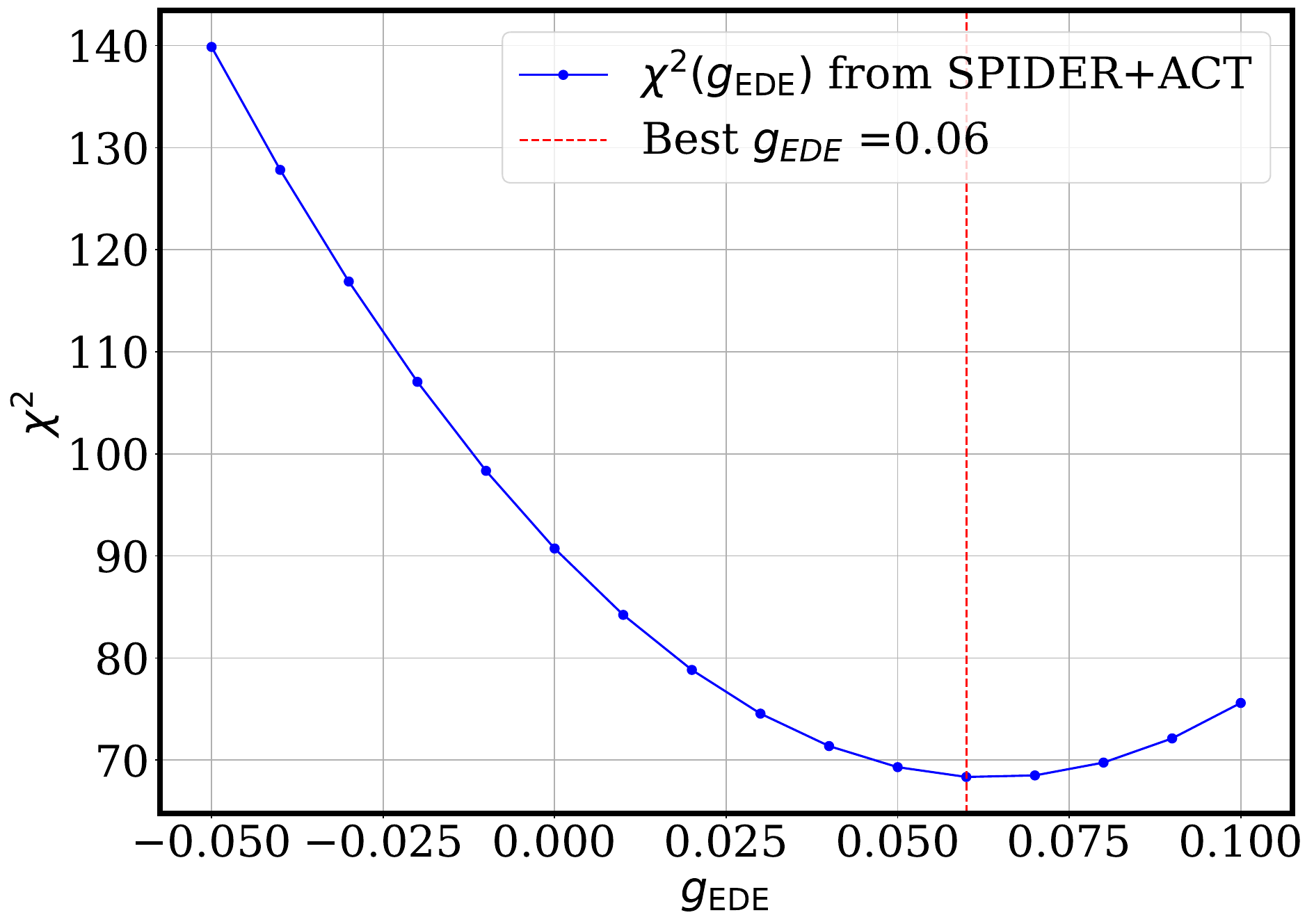} \\
        (d)   & (e)   & (f)  \\ 

        \includegraphics[width=0.3\linewidth]{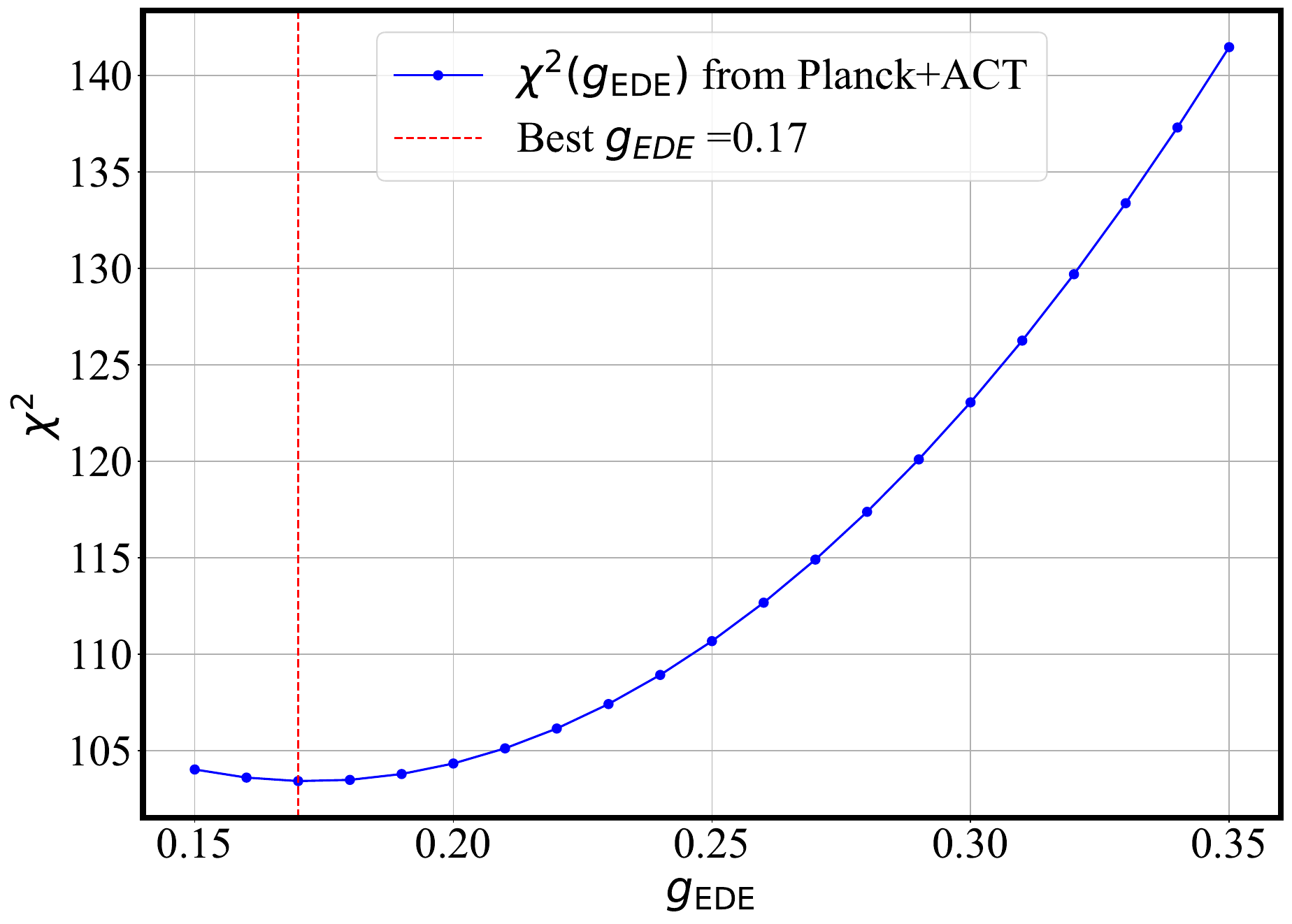} &
        \includegraphics[width=0.31\linewidth]{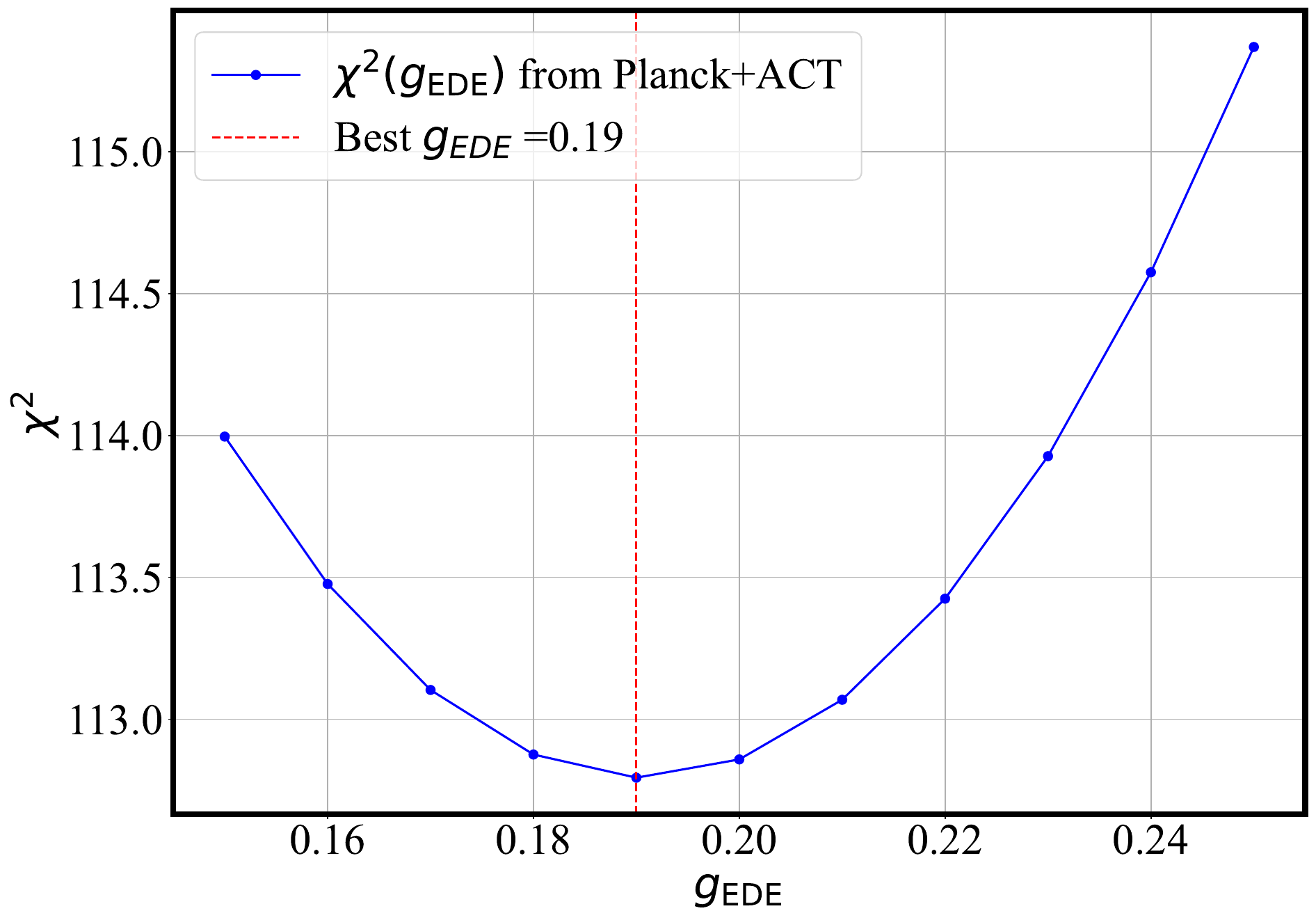} &
        \includegraphics[width=0.3\linewidth]{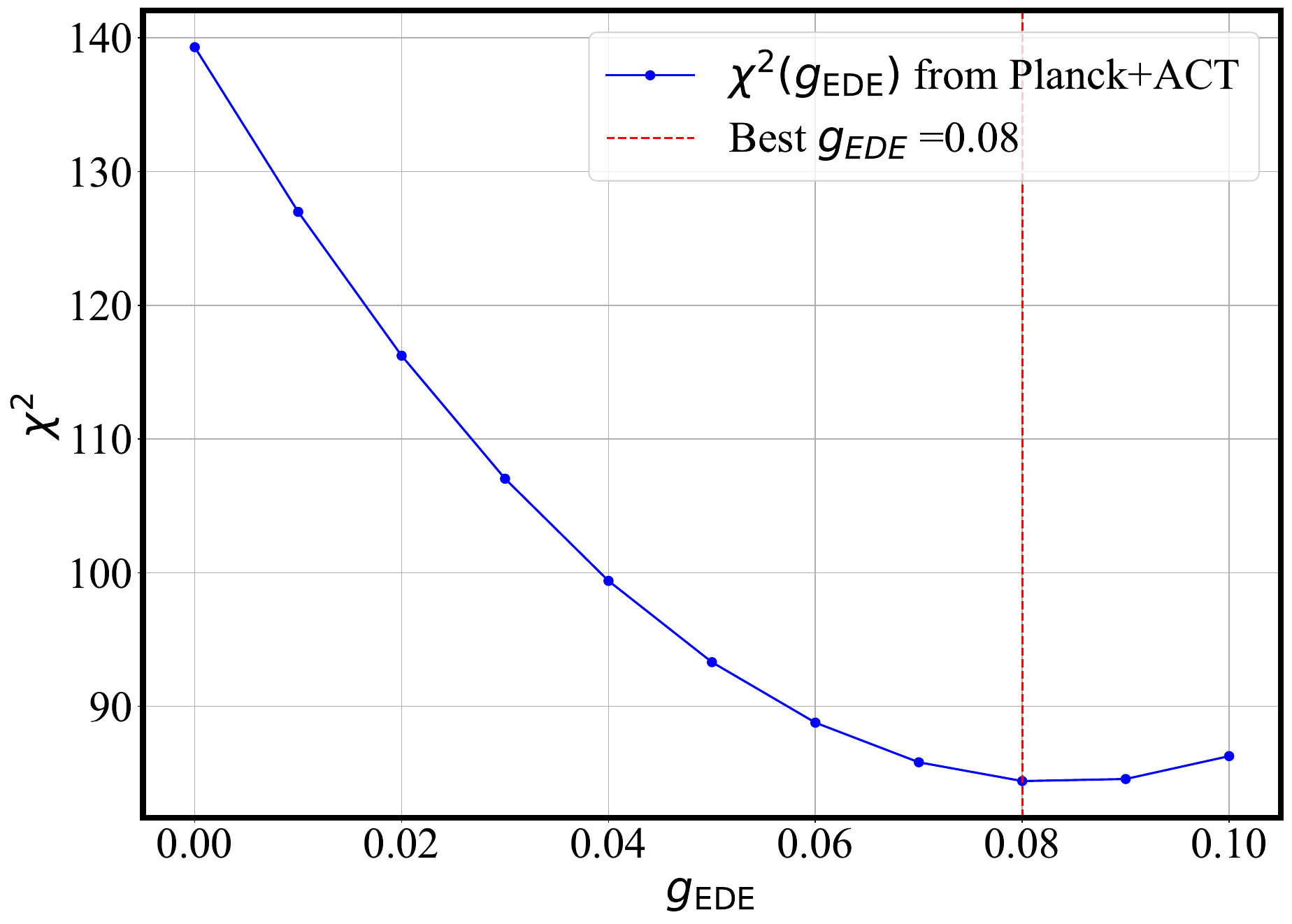} \\
        (g)   & (h)   & (i)  \\ 
    \end{tabular}
    \caption{\label{fig:12} {The $\chi^2$ and best-fit of $g_{EDE}$ (the coupling constant $gM_{Pl}$ in EDE model) results from SPIDER+Planck, SPIDER+ACT, Planck+ACT $EB$ dataset, respectively. The first, second, and third columns represent the nine fundamental parameters derived from Base, Base+SH0ES, and BSL, respectively. }}
\end{figure}

{In this appendix, we present the $\chi^2$ values and the best-fit constraints on the coupling constant $g_{\mathrm{EDE}}$ (defined as $g M_{\mathrm{Pl}}$ in the EDE model). The results obtained from the SPIDER, Planck, ACT, and SPIDER+Planck+ACT data sets are shown in Fig. \ref{fig:11}. The first, second, and third columns correspond to cosmological parameters derived from the Base, Base+SH0ES, and BSL datasets, respectively. 
Fig. \ref{fig:12} shows the constraints on $g_{\mathrm{EDE}}$ and the corresponding $\chi^2$ values obtained from the combined data of SPIDER+Planck, SPIDER+ACT, and Planck+ACT. Similarly, the first, second, and third columns correspond to cosmological parameters obtained from the Base, Base+SH0ES, and BSL cases, respectively. }

\begin{thebibliography}{99}

\bibitem{Komatsu:2014ioa}
E.~Komatsu \textit{et al.} [WMAP Science Team],
PTEP \textbf{2014}, 06B102 (2014)
doi:10.1093/ptep/ptu083
[arXiv:1404.5415 [astro-ph.CO]].

\bibitem{Planck:2018vyg}
N.~Aghanim \textit{et al.} [Planck],
Astron. Astrophys. \textbf{641}, A6 (2020)
[erratum: Astron. Astrophys. \textbf{652}, C4 (2021)]
doi:10.1051/0004-6361/201833910
[arXiv:1807.06209 [astro-ph.CO]].

\bibitem{SPT-3G:2021eoc}
D.~Dutcher \textit{et al.} [SPT-3G],
Phys. Rev. D \textbf{104}, no.2, 022003 (2021)
doi:10.1103/PhysRevD.104.022003
[arXiv:2101.01684 [astro-ph.CO]].

\bibitem{BICEP:2021xfz}
P.~A.~R.~Ade \textit{et al.} [BICEP and Keck],
Phys. Rev. Lett. \textbf{127}, no.15, 151301 (2021)
doi:10.1103/PhysRevLett.127.151301
[arXiv:2110.00483 [astro-ph.CO]].

\bibitem{SPIDER:2021ncy}
P.~A.~R.~Ade \textit{et al.} [SPIDER],
Astrophys. J. \textbf{927}, no.2, 174 (2022)
doi:10.3847/1538-4357/ac20df
[arXiv:2103.13334 [astro-ph.CO]].

\bibitem{Abdalla:2022yfr}
E.~Abdalla, G.~Franco Abell{\'a}n, A.~Aboubrahim, A.~Agnello, O.~Akarsu, Y.~Akrami, G.~Alestas, D.~Aloni, L.~Amendola and L.~A.~Anchordoqui, \textit{et al.}
JHEAp \textbf{34}, 49-211 (2022)
doi:10.1016/j.jheap.2022.04.002
[arXiv:2203.06142 [astro-ph.CO]].

\bibitem{Feng:2004mq}
B.~Feng, H.~Li, M.~z.~Li and X.~m.~Zhang,
Phys. Lett. B \textbf{620}, 27-32 (2005)
doi:10.1016/j.physletb.2005.06.009
[arXiv:hep-ph/0406269 [hep-ph]].

\bibitem{Komatsu:2022nvu}
E.~Komatsu,
Nature Rev. Phys. \textbf{4}, no.7, 452-469 (2022)
doi:10.1038/s42254-022-00452-4
[arXiv:2202.13919 [astro-ph.CO]].

\bibitem{Bernal:2016gxb}
J.~L.~Bernal, L.~Verde and A.~G.~Riess,
JCAP \textbf{10}, 019 (2016)
doi:10.1088/1475-7516/2016/10/019
[arXiv:1607.05617 [astro-ph.CO]].

\bibitem{Carroll:1989vb}
S.~M.~Carroll, G.~B.~Field and R.~Jackiw,
Phys. Rev. D \textbf{41}, 1231 (1990)
doi:10.1103/PhysRevD.41.1231

\bibitem{Carroll:1991zs}
S.~M.~Carroll and G.~B.~Field,
Phys. Rev. D \textbf{43}, 3789 (1991)
doi:10.1103/PhysRevD.43.3789

\bibitem{Harari:1992ea}
D.~Harari and P.~Sikivie,
Phys. Lett. B \textbf{289}, 67-72 (1992)
doi:10.1016/0370-2693(92)91363-E

\bibitem{Minami:2020odp}
Y.~Minami and E.~Komatsu,
Phys. Rev. Lett. \textbf{125}, no.22, 221301 (2020)
doi:10.1103/PhysRevLett.125.221301
[arXiv:2011.11254 [astro-ph.CO]].

\bibitem{Eskilt:2022cff}
J.~R.~Eskilt and E.~Komatsu,
Phys. Rev. D \textbf{106}, no.6, 063503 (2022)
doi:10.1103/PhysRevD.106.063503
[arXiv:2205.13962 [astro-ph.CO]].

\bibitem{Lue:1998mq}
A.~Lue, L.~M.~Wang and M.~Kamionkowski,
Phys. Rev. Lett. \textbf{83}, 1506-1509 (1999)
doi:10.1103/PhysRevLett.83.1506
[arXiv:astro-ph/9812088 [astro-ph]].

\bibitem{Naokawa:2023upt}
F.~Naokawa and T.~Namikawa,
Phys. Rev. D \textbf{108}, no.6, 063525 (2023)
doi:10.1103/PhysRevD.108.063525
[arXiv:2305.13976 [astro-ph.CO]].

\bibitem{Namikawa:2023zux}
T.~Namikawa and I.~Obata,
Phys. Rev. D \textbf{108}, no.8, 8 (2023)
doi:10.1103/PhysRevD.108.083510
[arXiv:2306.08875 [astro-ph.CO]].

\bibitem{Ferreira:2023jbu}
R.~Z.~Ferreira, S.~Gasparotto, T.~Hiramatsu, I.~Obata and O.~Pujolas,
JCAP \textbf{05}, 066 (2024)
doi:10.1088/1475-7516/2024/05/066
[arXiv:2312.14104 [hep-ph]].

\bibitem{Yin:2023srb}
L.~Yin, J.~Kochappan, T.~Ghosh and B.~H.~Lee,
JCAP \textbf{10}, 007 (2023)
doi:10.1088/1475-7516/2023/10/007
[arXiv:2305.07937 [astro-ph.CO]].

\bibitem{Greco:2024oie}
A.~Greco, N.~Bartolo and A.~Gruppuso,
JCAP \textbf{10}, 028 (2024)
doi:10.1088/1475-7516/2024/10/028
[arXiv:2401.07079 [astro-ph.CO]].

\bibitem{Namikawa:2024dgj}
T.~Namikawa,
Phys. Rev. D \textbf{109}, no.12, 12 (2024)
doi:10.1103/PhysRevD.109.123521
[arXiv:2404.13771 [astro-ph.CO]].


\bibitem{Sullivan:2025btc}
R.~M.~Sullivan, A.~Abghari, P.~Diego-Palazuelos, L.~T.~Hergt and D.~Scott,
JCAP \textbf{25}, no.06, 025 (2020)
doi:10.1088/1475-7516/2025/06/025
[arXiv:2502.07654 [astro-ph.CO]].

\bibitem{LiteBIRD:2025yfb}
E.~de la Hoz \textit{et al.} [LiteBIRD],
JCAP \textbf{07}, 083 (2025)
doi:10.1088/1475-7516/2025/07/083
[arXiv:2503.22322 [astro-ph.CO]].

\bibitem{Lonappan:2025hwz}
A.~I.~Lonappan, B.~Keating and K.~Arnold,
Phys. Rev. D \textbf{112}, no.2, 023555 (2025)
doi:10.1103/yxmh-rh9z
[arXiv:2504.13154 [astro-ph.CO]].

\bibitem{Namikawa:2025sft}
T.~Namikawa, K.~Murai and F.~Naokawa,
[arXiv:2506.20824 [astro-ph.CO]].

\bibitem{Ballardini:2025apf}
M.~Ballardini, A.~Gruppuso, S.~Paradiso, S.~S.~Sirletti and P.~Natoli,
JCAP \textbf{09}, 075 (2025)
doi:10.1088/1475-7516/2025/09/075
[arXiv:2507.16714 [astro-ph.CO]].

\bibitem{Namikawa:2025doa}
T.~Namikawa,
Phys. Rev. Lett. \textbf{135}, no.16, 161004 (2025)
doi:10.1103/qgnn-6hsf
[arXiv:2506.22999 [astro-ph.CO]].

\bibitem{Diego-Palazuelos:2025dmh}
P.~Diego-Palazuelos and E.~Komatsu,
[arXiv:2509.13654 [astro-ph.CO]].

\bibitem{Feng:2006dp}
B.~Feng, M.~Li, J.~Q.~Xia, X.~Chen and X.~Zhang,
Phys. Rev. Lett. \textbf{96}, 221302 (2006)
doi:10.1103/PhysRevLett.96.221302
[arXiv:astro-ph/0601095 [astro-ph]].

\bibitem{Li:2006ss}
M.~Li, J.~Q.~Xia, H.~Li and X.~Zhang,
Phys. Lett. B \textbf{651}, 357-362 (2007)
doi:10.1016/j.physletb.2007.06.050
[arXiv:hep-ph/0611192 [hep-ph]].

\bibitem{Li:2008tma}
M.~Li and X.~Zhang,
Phys. Rev. D \textbf{78}, 103516 (2008)
doi:10.1103/PhysRevD.78.103516
[arXiv:0810.0403 [astro-ph]].

\bibitem{Xia:2008si}
J.~Q.~Xia, H.~Li, G.~B.~Zhao and X.~Zhang,
Astrophys. J. Lett. \textbf{679}, L61-L63 (2008)
doi:10.1086/589447
[arXiv:0803.2350 [astro-ph]].

\bibitem{Xia:2007qs}
J.~Q.~Xia, H.~Li, X.~l.~Wang and X.~m.~Zhang,
Astron. Astrophys. \textbf{483}, 715-718 (2008)
doi:10.1051/0004-6361:200809410
[arXiv:0710.3325 [hep-ph]].

\bibitem{Xia:2009ah}
J.~Q.~Xia, H.~Li and X.~Zhang,
Phys. Lett. B \textbf{687}, 129-132 (2010)
doi:10.1016/j.physletb.2010.03.038
[arXiv:0908.1876 [astro-ph.CO]].

\bibitem{Li:2014oia}
S.~Y.~Li, J.~Q.~Xia, M.~Li, H.~Li and X.~Zhang,
Astrophys. J. \textbf{799}, no.2, 211 (2015)
doi:10.1088/0004-637X/799/2/211
[arXiv:1405.5637 [astro-ph.CO]].

\bibitem{Li:2013vga}
M.~Li and B.~Yu,
JCAP \textbf{06}, 016 (2013)
doi:10.1088/1475-7516/2013/06/016
[arXiv:1303.1881 [astro-ph.CO]].

\bibitem{Luongo:2021nqh}
O.~Luongo, M.~Muccino, E.~{\'O}.~Colg{\'a}in, M.~M.~Sheikh-Jabbari and L.~Yin,
Phys. Rev. D \textbf{105}, no.10, 103510 (2022)
doi:10.1103/PhysRevD.105.103510
[arXiv:2108.13228 [astro-ph.CO]].

\bibitem{Krishnan:2021dyb}
C.~Krishnan, R.~Mohayaee, E.~{\'O}.~Colg{\'a}in, M.~M.~Sheikh-Jabbari and L.~Yin,
Class. Quant. Grav. \textbf{38}, no.18, 184001 (2021)
doi:10.1088/1361-6382/ac1a81
[arXiv:2105.09790 [astro-ph.CO]].

\bibitem{Namikawa:2024sax}
T.~Namikawa,
Phys. Rev. D \textbf{111}, no.2, 2 (2025)
doi:10.1103/PhysRevD.111.023501
[arXiv:2410.05149 [astro-ph.CO]].
 
\bibitem{Choi:2021aze}
G.~Choi, W.~Lin, L.~Visinelli and T.~T.~Yanagida,
Phys. Rev. D \textbf{104}, no.10, L101302 (2021)
doi:10.1103/PhysRevD.104.L101302
[arXiv:2106.12602 [hep-ph]].

\bibitem{Nakatsuka:2022epj}
H.~Nakatsuka, T.~Namikawa and E.~Komatsu,
Phys. Rev. D \textbf{105}, no.12, 123509 (2022)
doi:10.1103/PhysRevD.105.123509
[arXiv:2203.08560 [astro-ph.CO]].

\bibitem{Murai:2022zur}
K.~Murai, F.~Naokawa, T.~Namikawa and E.~Komatsu,
Phys. Rev. D \textbf{107}, no.4, L041302 (2023)
doi:10.1103/PhysRevD.107.L041302
[arXiv:2209.07804 [astro-ph.CO]].

\bibitem{POLARBEAR:2019snn}
S.~Adachi \textit{et al.} [POLARBEAR],
Phys. Rev. Lett. \textbf{124}, no.13, 131301 (2020)
doi:10.1103/PhysRevLett.124.131301
[arXiv:1909.13832 [astro-ph.CO]].

\bibitem{ACT:2025fju}
T.~Louis \textit{et al.} [ACT],
[arXiv:2503.14452 [astro-ph.CO]].

\bibitem{SimonsObservatory:2018koc}
P.~Ade \textit{et al.} [Simons Observatory],
JCAP \textbf{02}, 056 (2019)
doi:10.1088/1475-7516/2019/02/056
[arXiv:1808.07445 [astro-ph.CO]].

\bibitem{Gao:2017cra}
H.~Gao, C.~Liu, Z.~Li, Y.~Liu, Y.~Li, S.~Li, H.~Li, G.~Gao, F.~Lu and X.~Zhang,
Radiat. Detect. Technol. Methods \textbf{1}, no.2, 12 (2017)
doi:10.1007/s41605-017-0013-3

\bibitem{Li:2017drr}
H.~Li, S.~Y.~Li, Y.~Liu, Y.~P.~Li, Y.~Cai, M.~Li, G.~B.~Zhao, C.~Z.~Liu, Z.~W.~Li and H.~Xu, \textit{et al.}
Natl. Sci. Rev. \textbf{6}, no.1, 145-154 (2019)
doi:10.1093/nsr/nwy019
[arXiv:1710.03047 [astro-ph.CO]].

\bibitem{LiteBIRD:2022cnt}
E.~Allys \textit{et al.} [LiteBIRD],
PTEP \textbf{2023}, no.4, 042F01 (2023)
doi:10.1093/ptep/ptac150
[arXiv:2202.02773 [astro-ph.IM]].

\bibitem{Riess:2021jrx}
A.~G.~Riess, W.~Yuan, L.~M.~Macri, D.~Scolnic, D.~Brout, S.~Casertano, D.~O.~Jones, Y.~Murakami, L.~Breuval and T.~G.~Brink, \textit{et al.}
Astrophys. J. Lett. \textbf{934}, no.1, L7 (2022)
doi:10.3847/2041-8213/ac5c5b
[arXiv:2112.04510 [astro-ph.CO]].

\bibitem{Escamilla:2024ahl}
L.~A.~Escamilla, E.~{\"O}z{\"u}lker, {\"O}.~Akarsu, E.~Di Valentino and J.~A.~V{\'a}zquez,
Mon. Not. Roy. Astron. Soc. \textbf{836}, 854 (2025)
doi:10.1093/mnras/staf1732
[arXiv:2408.12516 [astro-ph.CO]].

\bibitem{Du:2024pai}
G.~H.~Du, P.~J.~Wu, T.~N.~Li and X.~Zhang,
Eur. Phys. J. C \textbf{85}, no.4, 392 (2025)
doi:10.1140/epjc/s10052-025-14094-0
[arXiv:2407.15640 [astro-ph.CO]].

\bibitem{Li:2024qso}
T.~N.~Li, P.~J.~Wu, G.~H.~Du, S.~J.~Jin, H.~L.~Li, J.~F.~Zhang and X.~Zhang,
Astrophys. J. \textbf{976}, no.1, 1 (2024)
doi:10.3847/1538-4357/ad87f0
[arXiv:2407.14934 [astro-ph.CO]].

\bibitem{Cai:2025mas}
Y.~Cai, X.~Ren, T.~Qiu, M.~Li and X.~Zhang,
[arXiv:2505.24732 [astro-ph.CO]].

\bibitem{Li:2025nnk}
H.~H.~Li, X.~z.~Zhang, T.~Qiu and J.~Q.~Xia,
JCAP \textbf{07}, 056 (2025)
doi:10.1088/1475-7516/2025/07/056
[arXiv:2503.06941 [astro-ph.CO]].

\bibitem{Qiu:2024sdd}
T.~Qiu and M.~Zhu,
Phys. Rev. D \textbf{111}, no.4, 043508 (2025)
doi:10.1103/PhysRevD.111.043508
[arXiv:2408.06582 [gr-qc]].

\bibitem{Feng:2025mlo}
L.~Feng, T.~N.~Li, G.~H.~Du, J.~F.~Zhang and X.~Zhang,
Phys. Dark Univ. \textbf{48}, 101935 (2025)
doi:10.1016/j.dark.2025.101935
[arXiv:2503.10423 [astro-ph.CO]].

\bibitem{Smith:2025icl}
A.~Smith, E.~{\"O}z{\"u}lker, E.~Di Valentino and C.~van de Bruck,
[arXiv:2510.21931 [astro-ph.CO]].

\bibitem{Lee:2025yvn}
J.~Lee, K.~Murai, F.~Takahashi and W.~Yin,
Phys. Rev. D \textbf{112}, no.4, 043538 (2025)
doi:10.1103/pjf4-sn4v
[arXiv:2503.18417 [hep-ph]].

\bibitem{Piras:2025eip}
D.~Piras, L.~Herold, L.~Lucie-Smith and E.~Komatsu,
Phys. Rev. D \textbf{111}, no.8, 083537 (2025)
doi:10.1103/PhysRevD.111.083537
[arXiv:2502.09810 [astro-ph.CO]].

\bibitem{Li:2025owk}
T.~N.~Li, G.~H.~Du, Y.~H.~Li, P.~J.~Wu, S.~J.~Jin, J.~F.~Zhang and X.~Zhang,
Sci. China Phys. Mech. Astron. \textbf{69}, no.1, 210413 (2026)
doi:10.1007/s11433-025-2771-5
[arXiv:2501.07361 [astro-ph.CO]].

\bibitem{Yin:2020dwl}
L.~Yin,
Eur. Phys. J. C \textbf{82}, no.1, 78 (2022)
doi:10.1140/epjc/s10052-022-10020-w
[arXiv:2012.13917 [astro-ph.CO]].

\bibitem{Colgain:2021pmf}
E.~{\'O}.~Colg{\'a}in, M.~M.~Sheikh-Jabbari and L.~Yin,
Phys. Rev. D \textbf{104}, no.2, 023510 (2021)
doi:10.1103/PhysRevD.104.023510
[arXiv:2104.01930 [astro-ph.CO]].

\bibitem{Krishnan:2021jmh}
C.~Krishnan, R.~Mohayaee, E.~{\'O}.~Colg{\'a}in, M.~M.~Sheikh-Jabbari and L.~Yin,
Phys. Rev. D \textbf{105}, no.6, 063514 (2022)
doi:10.1103/PhysRevD.105.063514
[arXiv:2106.02532 [astro-ph.CO]].

\bibitem{Akarsu:2022lhx}
O.~Akarsu, E.~O.~Colgain, E.~{\"O}zulker, S.~Thakur and L.~Yin,
Phys. Rev. D \textbf{107}, no.12, 123526 (2023)
doi:10.1103/PhysRevD.107.123526
[arXiv:2207.10609 [astro-ph.CO]].

\bibitem{Poulin:2018cxd}
V.~Poulin, T.~L.~Smith, T.~Karwal and M.~Kamionkowski,
Phys. Rev. Lett. \textbf{122}, no.22, 221301 (2019)
doi:10.1103/PhysRevLett.122.221301
[arXiv:1811.04083 [astro-ph.CO]].

\bibitem{Herold:2023vzx}
L.~Herold,
doi:10.5282/edoc.32098

\bibitem{Efstathiou:2023fbn}
G.~Efstathiou, E.~Rosenberg and V.~Poulin,
Phys. Rev. Lett. \textbf{132}, no.22, 221002 (2024)
doi:10.1103/PhysRevLett.132.221002
[arXiv:2311.00524 [astro-ph.CO]].

\bibitem{Simon:2024jmu}
T.~Simon, T.~Adi, J.~L.~Bernal, E.~D.~Kovetz, V.~Poulin and T.~L.~Smith,
Phys. Rev. D \textbf{111}, no.2, 2 (2025)
doi:10.1103/PhysRevD.111.023523
[arXiv:2410.21459 [astro-ph.CO]].

\bibitem{Lin:2025gne}
W.~Lin, L.~Visinelli and T.~T.~Yanagida,
JCAP \textbf{10}, 023 (2025)
doi:10.1088/1475-7516/2025/10/023
[arXiv:2504.17638 [astro-ph.CO]].

\bibitem{Eskilt:2023nxm}
J.~R.~Eskilt, L.~Herold, E.~Komatsu, K.~Murai, T.~Namikawa and F.~Naokawa,
Phys. Rev. Lett. \textbf{131}, no.12, 121001 (2023)
doi:10.1103/PhysRevLett.131.121001
[arXiv:2303.15369 [astro-ph.CO]].

\bibitem{Kochappan:2024jyf}
J.~Kochappan, L.~Yin, B.~H.~Lee and T.~Ghosh,
Phys. Rev. D \textbf{112}, no.6, 063562 (2025)
doi:10.1103/x1qj-t4jz
[arXiv:2408.09521 [astro-ph.CO]].

\bibitem{Caldwell:2003vp}
R.~R.~Caldwell, M.~Doran, C.~M.~Mueller, G.~Schafer and C.~Wetterich,
Astrophys. J. Lett. \textbf{591}, L75-L78 (2003)
doi:10.1086/376975
[arXiv:astro-ph/0302505 [astro-ph]].

\bibitem{Smith:2019ihp}
T.~L.~Smith, V.~Poulin and M.~A.~Amin,
Phys. Rev. D \textbf{101}, no.6, 063523 (2020)
doi:10.1103/PhysRevD.101.063523
[arXiv:1908.06995 [astro-ph.CO]].

\bibitem{Berghaus:2019cls}
K.~V.~Berghaus and T.~Karwal,
Phys. Rev. D \textbf{101}, no.8, 083537 (2020)
doi:10.1103/PhysRevD.101.083537
[arXiv:1911.06281 [astro-ph.CO]].

\bibitem{Alexander:2019rsc}
S.~Alexander and E.~McDonough,
Phys. Lett. B \textbf{797}, 134830 (2019)
doi:10.1016/j.physletb.2019.134830
[arXiv:1904.08912 [astro-ph.CO]].

\bibitem{Chudaykin:2020acu}
A.~Chudaykin, D.~Gorbunov and N.~Nedelko,
JCAP \textbf{08}, 013 (2020)
doi:10.1088/1475-7516/2020/08/013
[arXiv:2004.13046 [astro-ph.CO]].

\bibitem{Agrawal:2019lmo}
P.~Agrawal, F.~Y.~Cyr-Racine, D.~Pinner and L.~Randall,
Phys. Dark Univ. \textbf{42}, 101347 (2023)
doi:10.1016/j.dark.2023.101347
[arXiv:1904.01016 [astro-ph.CO]].

\bibitem{Niedermann:2019olb}
F.~Niedermann and M.~S.~Sloth,
Phys. Rev. D \textbf{103}, no.4, L041303 (2021)
doi:10.1103/PhysRevD.103.L041303
[arXiv:1910.10739 [astro-ph.CO]].

\bibitem{Freese:2004vs}
K.~Freese and D.~Spolyar,
JCAP \textbf{07}, 007 (2005)
doi:10.1088/1475-7516/2005/07/007
[arXiv:hep-ph/0412145 [hep-ph]].

\bibitem{Ye:2020btb}
G.~Ye and Y.~S.~Piao,
Phys. Rev. D \textbf{101}, no.8, 083507 (2020)
doi:10.1103/PhysRevD.101.083507
[arXiv:2001.02451 [astro-ph.CO]].

\bibitem{Akarsu:2019hmw}
{\"O}.~Akarsu, J.~D.~Barrow, L.~A.~Escamilla and J.~A.~Vazquez,
Phys. Rev. D \textbf{101}, no.6, 063528 (2020)
doi:10.1103/PhysRevD.101.063528
[arXiv:1912.08751 [astro-ph.CO]].

\bibitem{Lin:2019qug}
M.~X.~Lin, G.~Benevento, W.~Hu and M.~Raveri,
Phys. Rev. D \textbf{100}, no.6, 063542 (2019)
doi:10.1103/PhysRevD.100.063542
[arXiv:1905.12618 [astro-ph.CO]].

\bibitem{Braglia:2020bym}
M.~Braglia, W.~T.~Emond, F.~Finelli, A.~E.~Gumrukcuoglu and K.~Koyama,
Phys. Rev. D \textbf{102}, no.8, 083513 (2020)
doi:10.1103/PhysRevD.102.083513
[arXiv:2005.14053 [astro-ph.CO]].

\bibitem{Finelli:2008jv}
F.~Finelli and M.~Galaverni,
Phys. Rev. D \textbf{79}, 063002 (2009)
doi:10.1103/PhysRevD.79.063002
[arXiv:0802.4210 [astro-ph]].

\bibitem{Galaverni:2023zhv}
M.~Galaverni, F.~Finelli and D.~Paoletti,
Phys. Rev. D \textbf{107}, no.8, 083529 (2023)
doi:10.1103/PhysRevD.107.083529
[arXiv:2301.07971 [astro-ph.CO]].

\bibitem{Hill:2020osr}
J.~C.~Hill, E.~McDonough, M.~W.~Toomey and S.~Alexander,
Phys. Rev. D \textbf{102}, no.4, 043507 (2020)
doi:10.1103/PhysRevD.102.043507
[arXiv:2003.07355 [astro-ph.CO]].

\bibitem{Lesgourgues:2011re}
J.~Lesgourgues,
[arXiv:1104.2932 [astro-ph.IM]].

\bibitem{Blas:2011rf}
D.~Blas, J.~Lesgourgues and T.~Tram,
JCAP \textbf{07}, 034 (2011)
doi:10.1088/1475-7516/2011/07/034
[arXiv:1104.2933 [astro-ph.CO]].

\bibitem{Planck:2019nip}
N.~Aghanim \textit{et al.} [Planck],
Astron. Astrophys. \textbf{641}, A5 (2020)
doi:10.1051/0004-6361/201936386
[arXiv:1907.12875 [astro-ph.CO]].

\bibitem{BOSS:2016wmc}
S.~Alam \textit{et al.} [BOSS],
Mon. Not. Roy. Astron. Soc. \textbf{470}, no.3, 2617-2652 (2017)
doi:10.1093/mnras/stx721
[arXiv:1607.03155 [astro-ph.CO]].


\end{thebibliography}
\end{document}